\documentclass[11pt,a4paper]{article}
\usepackage[T1]{fontenc}
\usepackage[utf8]{inputenc}
\usepackage{authblk}
\usepackage{longtable}
\usepackage{url}
\usepackage[ruled]{algorithm2e}
\usepackage{multirow} 
\usepackage{wasysym }
\usepackage{graphicx}
\usepackage[margin=0.8in]{geometry}

\title{Friend or Foe? Fake Profile Identification in Online Social Networks}
\author{Michael Fire, Dima Kagan, Aviad Elyashar, and Yuval Elovici\thanks{Email: \{mickyfi, kagandi,aviade, elovicig@bgu.ac.il\}}}

\affil{Telekom Innovation Laboratories  at Ben-Gurion University of the Negev \\
Department of Information Systems Engineering, Ben Gurion University}

\date{}

\begin{document}
\maketitle


\begin{abstract}
The amount of personal information unwillingly exposed by users on online social networks is staggering, as shown in recent research.  Moreover, recent reports indicate that these networks are infested with tens of millions of fake users profiles, which may jeopardize the users' security and privacy. To identify fake users in such networks and to improve users' security and privacy, we developed the Social Privacy Protector software for Facebook. This software contains three protection layers, which improve user privacy by implementing different methods. The software first identifies a user's friends who might pose a threat and then restricts this ``friend's'' exposure to the user's personal information. The second layer is an expansion of Facebook's basic privacy settings based on different types of social network usage profiles. The third layer alerts users about the number of installed applications on their Facebook profile, which have access to their private information. An initial version of the  Social Privacy Protection software received high media coverage, and more than 3,000 users from more than twenty countries have installed the software, out of which 527 used the software to restrict more than nine thousand friends. In addition, we estimate that more than a hundred users accepted the software's recommendations and removed at least 1,792 Facebook applications from their profiles. By analyzing the unique dataset obtained by the software in combination with machine learning techniques, we developed classifiers, which are able to predict which Facebook profiles have high probabilities of being fake and therefore, threaten the user's well-being. 
Moreover, in this study, we present statistics on users' privacy settings and statistics of the number of applications installed on Facebook profiles. Both statistics are obtained by the Social Privacy Protector software. These statistics alarmingly demonstrate how exposed Facebook users information is to both fake profile attacks and  third party Facebook applications. 

\end{abstract}

\section{Introduction}
In recent years, online social networks have grown rapidly and today offer individuals endless possibilities for publicly expressing themselves, communicating with friends, and sharing information with people across the world. 
A recent survey~\cite{pewinternet} estimated that 65\% of adult internet users use online social network sites, such as Twitter~\cite{twitter}, LinkedIn~\cite{linkedin}, Google+~\cite{googleplus}, and Facebook~\cite{facebook}. 
As of October 2012, the Facebook social network has more than one billion active users monthly~\cite{facebook-newsroom}. 
On average, Facebook users have 138 friends and upload more than 219 billion pictures onto Facebook~\cite{facebook-newsroom}. 
Moreover, according to the Nielsen ``Social Media Report''~\cite{nielsen}, American Internet users spent more than 53.5 billion minutes on Facebook in the month of May 2011, making Facebook the leading web-brand in the United-States.

Due to the friendly nature of Facebook, users tend to disclose many personal details about themselves and about their connections.  These details can include date of birth, personal pictures, work place, email address, high school name, relationship statuses, and even phone numbers. Moreover, Bosmaf et al.~\cite{boshmaf}, discovered that an average of 80\% of studied Facebook users accepted friend requests from people they did not know if they shared more than 11 mutual friends.  
In many cases, accepting friend requests from strangers may result in the exposure of a user's personal information to third parties. 
Additionally, personal information of Facebook users can be exposed to third party Facebook applications~\cite{egele2011pox}. 
Another privacy concern deals with existing privacy settings,  for which the majority of Facebook users do not  match the security expectations~\cite{liu2011analyzing}.
 These results indicate that many users accidently or unknowingly publish private information leaving them more exposed than they assumed.  

If a user's personal information is disclosed to a malicious third party, it can be used to threaten the well-being of the user both online and in the real world. For example, a malicious user can use the gained personal information and send customized spam messages to a user in an attempt to lure such users onto malicious websites~\cite{stringhini}, or blackmail them into transferring money to the attacker's account~\cite{nelson}. To cover their tracks, social network attackers may use fake profiles. In fact, the number of fake profiles on Facebook can be counted in the tens of millions.
According to a recent report~\cite{facebook_fake}, Facebook estimates that 8.7\% (83.09 million), of its accounts do not belong to real profiles. Moreover, Facebook also estimates that 1.5\% (14.32 million), of its accounts are ``undesirable accounts'', which belong to users who may deliberately spread undesirable content, such as spam messages and malicious links, and threaten the security and privacy of other Facebook users.

In this study, we present the Social Privacy Protector software for protecting user privacy on Facebook (otherwise referred to as SPP). 
The SPP software consists of two main parts, namely, a Firefox add-on and a Facebook application.
The two parts provide Facebook users with three different layers of protection. 
The first layer, which is part of the Firefox add-on, enables Facebook users to easily control their profile privacy settings by simply choosing the most suitable profile privacy settings with just one click. 
The second layer, which is also part of the software Firefox add-on, notifies users of the number of applications installed on their profile which may impose a threat to their privacy. 
The third layer, a Facebook application, analyzes a user's friends list. By using simple heuristics (see Section~\ref{spp_features}), the application identifies which friends of a user are suspected as fake profiles and therefore impose a threat on the user's privacy. The application presents a convenient method for restricting the access of fake profiles to a user's personal information without removing them from the user's friends list.

At the end of June 2012, we launched an initial version of the SPP software as a ``free to use software''~\cite{spp,fire2012social} and received massive media coverage with hundreds of online articles and interviews in leading blogs and news websites, such as Fox news~\cite{foxnews} and NBC news~\cite{nbcnews}. 
Due to the media coverage, in less than four months, 3,017 users from more than twenty countries installed the SPP  Facebook application, 527 of which used the SPP Facebook application to restrict 9,005  friends\footnote{Due to the unexpected massive downloads and usage of the application our servers did not succeeded in supporting the large amount of users at once. Moreover, in our initial version, the SPP Facebook application did not support all the existing web browsers. Therefore, many users who installed the SPP software did not have the possibility to use it on demand.}. Moreover, at least 1,676 users installed the Firefox add-on out of which we estimate that 111 users used the add-on recommendation and removed more than 1,792 Facebook applications from their profiles (see Section~\ref{sec:results_app}). In addition, the  add-on also succeeded in collecting the Facebook privacy settings of 67 different Facebook users.

To our great surprise many of the SPP application users used the application to not only remove users that were recommended for removal, but to also manually search and restrict them by name, specific friends that have a higher likelihood of having profiles that belong to real people. 
The removal of real profiles also assists us in studying and constructing classifiers that identify real profiles recommended for restriction.
The collected data obtained from SPP users gave us a unique opportunity to learn more about user privacy on online social networks in general, and on Facebook in particular. 
Also, by using the unique data obtained from users restricting their Facebook friends, as well as by implementing machine learning techniques, we developed classifiers, which can identify user's friends that are recommended for restriction (see Section~\ref{sec:ml}). 
Our classifiers presented an AUC of up to 0.948, precision at 200 of up to 98\%, and an average users precision at 10  of up to 24\% (see Section~\ref{results}).  
Furthermore, these types of classifiers can also be used by online social network administrators to identify and remove fake profiles from the online social network.

In this study we also present statistics on Facebook user privacy settings, which were obtained by the SPP Add-on. These statistics demonstrate how exposed Facebook users' personal information is to fake profile attacks and third party applications (see Section~\ref{sec:results_addon}). 
For example, we show that out of 1,676 examined Facebook users, 10.68\% have more than a hundred Facebook applications installed on their profile,  and 30.31\% of the users have at least 40 Facebook applications installed. Moreover, out of 67 collected users' privacy settings the majority of the user's personal information is set up to be exposed to friends leaving the user's personal information exposed to fake friends. 

The remainder of this paper is organized as follows.
In Section~\ref{background}, we provide an overview of various related solutions, which better help protect the security and privacy of social network users. In addition, we also present an overview of similar studies, which used machine learning techniques to predict user properties, such as predicting users' links in social networks.
In Section~\ref{architecture}, we describe the SPP software architecture in detail.
In Section~\ref{sec:methods}, we describe the methods and experiments used in this study.
We describe the initial deployment of the SPP software and the methods used for the construction and evaluation of our machine learning classifiers.
In Section~\ref{results}, we present the results of our study, which include an evaluation of our classifiers and different users' privacy statistics obtained by the SPP software. 
In Section~\ref{sec:discussion}, we discuss the obtained results.
Lastly, in Section~\ref{sec:conclusions}, we present our conclusions from this study and offer future research directions.

\section{Related Work}
\label{background}
\subsection{Online Social Network Security and Privacy}
In recent years, due to the increasing number of privacy and security threats on online social network users, social network operators, security companies, and academic researchers have proposed various solutions to increase the security and privacy of social network users.

Social network operators attempt to better protect their users by adding authentication processes to ensure that a registered user represents a real live person~\cite{kuzma2011account}. 
Many social network operators, like Facebook, also offer their users a configurable user privacy setting that enables users to secure their personal data from other users in the network~\cite{liu2011analyzing,mahmood}.
Additional protection may include a shield against hackers, spammers, socialbots, identity cloning, phishing, and many other threats. 
For example, Facebook users have an option to report users in the network who harass other users in the network~\cite{facebook-report}. In addition, Facebook also developed and deployed an Immune System, which aims to protect its user from different online threats~\cite{stein2011facebook}.

Many commercial and open source products, such as Checkpoint's SocialGuard~\cite{zonealarm}, Websense's Defensio~\cite{defensio}, UnitedParents~\cite{unitedparents}, RecalimPrivacy~\cite{reclaimprivacy}, and PrivAware application~\cite{privaware}, offer online social network users tools for better protecting themselves. 
For example, the Websense's Defensio software aims to protect its users from spammers, adult content, and malicious scripts on Facebook.

In recent years, several published academic studies have proposed solutions for various social network threats. 
DeBarr and Wechsler~\cite{debarr} used the graph centrality measure to identify spammers. 
Wang~\cite{wang} presented techniques to classify spammers on Twitter based on content and graph features.  Stringhini et al.~\cite{stringhini} presented a solution for detecting spammers in social networks by using ``honey-profiles''. Egele et al.~\cite{egele2011pox} presented PoX, an extension for Facebook, which makes all requests for private data explicit to the user.
Yang et al.~\cite{yang2011uncovering} presented a method to identify fake profiles by analyzing different features, such as links' creation timestamps, and friend requests frequency.
Anwar and Fong~\cite{anwar2012visualization} presented the Reflective Policy Assessment tool, which aids users in examining their profiles from the viewpoint of another user in the network.
Rahman et al.~\cite{rahman2012efficient} presented the MyPageKeeper Facebook application,
which aims to protect Facebook users from damaging posts on the user's Facebook wall. 
In a later study~\cite{rahman2012frappe}, Rahman et al. also presented the FRAppE application for detecting malicious applications on Facebook. They discovered that 13\% of one hundred and eleven thousand Facebook applications in their dataset were malicious applications. 
Recently, Fire et al.~\cite{fire2012strangers} proposed a method for detecting fake profiles in online social networks based on anomalies in a fake user's social structure.

In this study, we present the the SPP software, which offers methods for improving Facebook user privacy\footnote{An initial version of the SPP software was described, as work in progress, in our previous paper~\cite{fire2012social}}. By using data collected by the SPP software and machine learning techniques, we present methods for constructing classifiers that can assist in identifying fake profiles. 

\subsection{Online Social Networks and Machine Learning}
With the increasing popularity of social networks many researchers had studied and used
a combination of data obtained from social networks and machine learning techniques to predict different user properties~\cite{liben,altshuler2012many,sakaki2010earthquake}. Furthermore, several studies used machine learning techniques  to improve user security in online social networks~\cite{stringhini,lee,fire2012strangers}. 
 
In 2007, Liben-Nowell and Kleinberg~\cite{liben} used machine learning techniques  to predict links between users in different social networks (also referred to as the link prediction problem). In 2010, Stringhini et al.~\cite{stringhini} proposed a method for detecting spammer profiles by using supervised learning algorithms. In the same year, Lee et al.~\cite{lee} used machine learning and honeypots to uncover spammers in MySpace and Twitter. Sakaki et al.~\cite{sakaki2010earthquake} used machine learning and content data analysis of Twitter users in order to detect events, such as earthquakes and typhoons, in real-time.
In 2012, Altshuler et al.~\cite{altshuler2012many} used machine learning techniques to predict different user's properties, such as origin and ethnicity, inside the ``Friends and Family'' social network, which was created by logs extracted from a user's mobile device. 
Recently, Fire et al.~\cite{fire2012strangers} used the online social network's topological features to identify fake users in different online social networks. 

As part of this study we present a method for recommending to a Facebook user which of his friends might be a fake profile and should therefore, be restricted. Our method is based on the connection properties between a Facebook user and its friends and by using supervised learning techniques. This type of problem is to some degree similar to the problem of predicting link strength, studied by Kahanda and Nevill~\cite{kahanda2009using}, and the problem of predicting positive and negative links (signed links), as   Leskovec et al.~\cite{leskovec2010predicting} studied. Similarly to the study held by Kahanda and Nevill, in this study we extract a different set of meta-content features, such as the number of pictures and videos both the user and his friends were tagged in. In this study, we also predict the type of a negative relationship between  users, similar to the study of Leskovec et al. However, in our study we aim to uncover fake profiles rather than identify the link sign or strength between two users. In addition, our study contrasts other studies, which used a major part of the social network topology to construct  classifiers~\cite{kahanda2009using,leskovec2010predicting,fire2012strangers}, because we construct our classifiers by using only  variations of the data collected in real-time from the user's point of view rather than data collected from the social network administrator's point of view. 
By using user data, which was obtained in real-time only, we were able to quickly analyze each user's friends list with fewer resources and without invading the user's friend's privacy.

\section{Social Privacy Protector Architecture}
\label{architecture}
\begin{figure}[ht]
\begin{center}

\includegraphics[
 width=0.8\textwidth,clip]{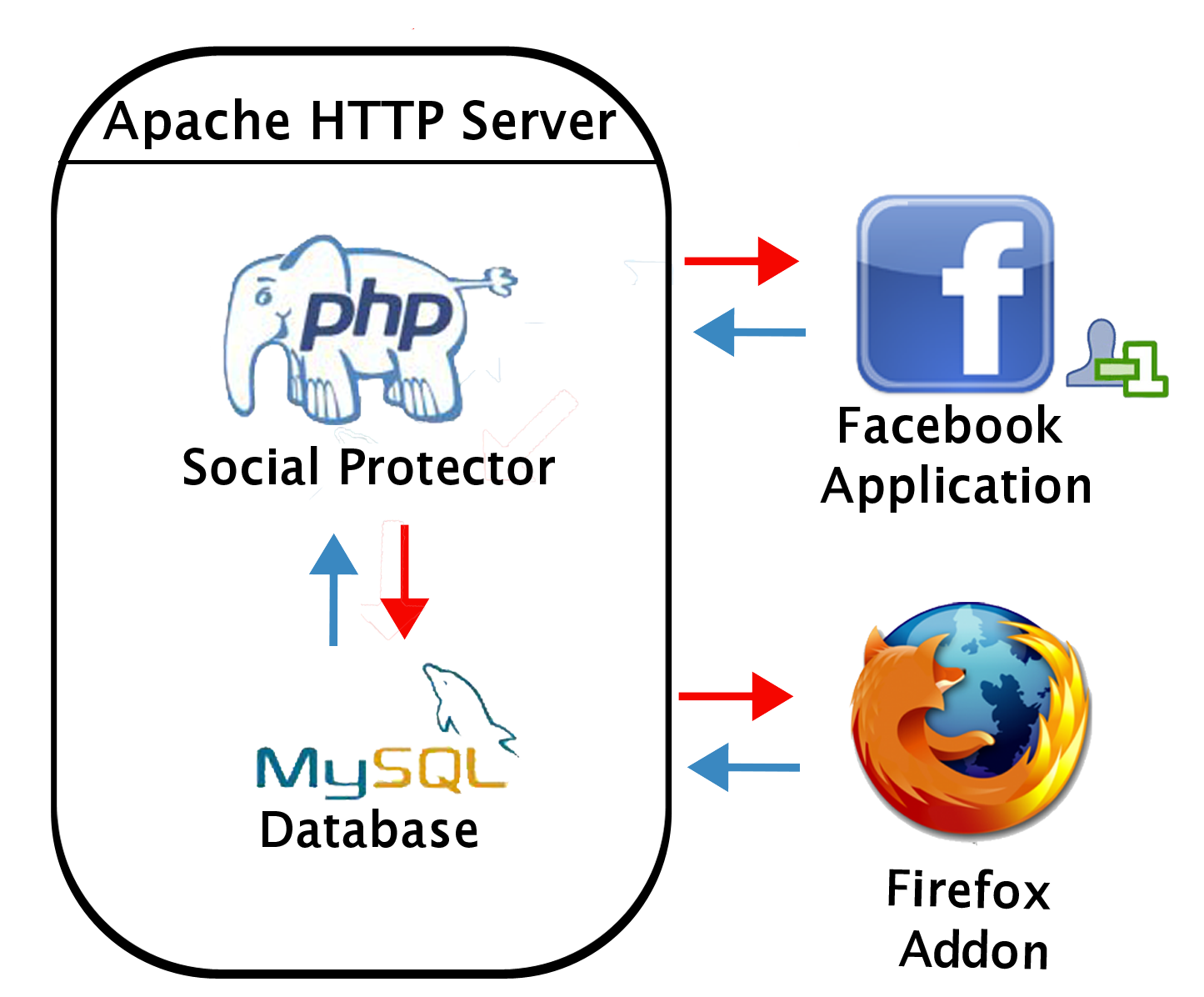}
\end{center}
\caption{Social Privacy Protector Architecture.}
\label{archit}
\end{figure}

To better protect the privacy of Facebook users we developed the \textit{Social Privacy Protector} software. The SPP software consists of three main parts (see Figure~\ref{archit}), which work in synergy: a) \textit{Friends Analyzer Facebook application} - which is responsible for identifying a user's friends who may pose a threat to the users privacy, b) \textit{SPP Firefox Add-on} - which analyzes the user's privacy settings and assists the user in improving privacy settings with just one click, and c) \textit{HTTP Server} - which is responsible for analyzing, storing, and caching software results for each user. 
In the remainder of this section, we describe in detail each individual part of the SPP software.

\subsection{The Friends Analyzer Facebook Application}
\begin{figure}[ht]
\begin{center}

\includegraphics[
width=0.8\textwidth,clip]{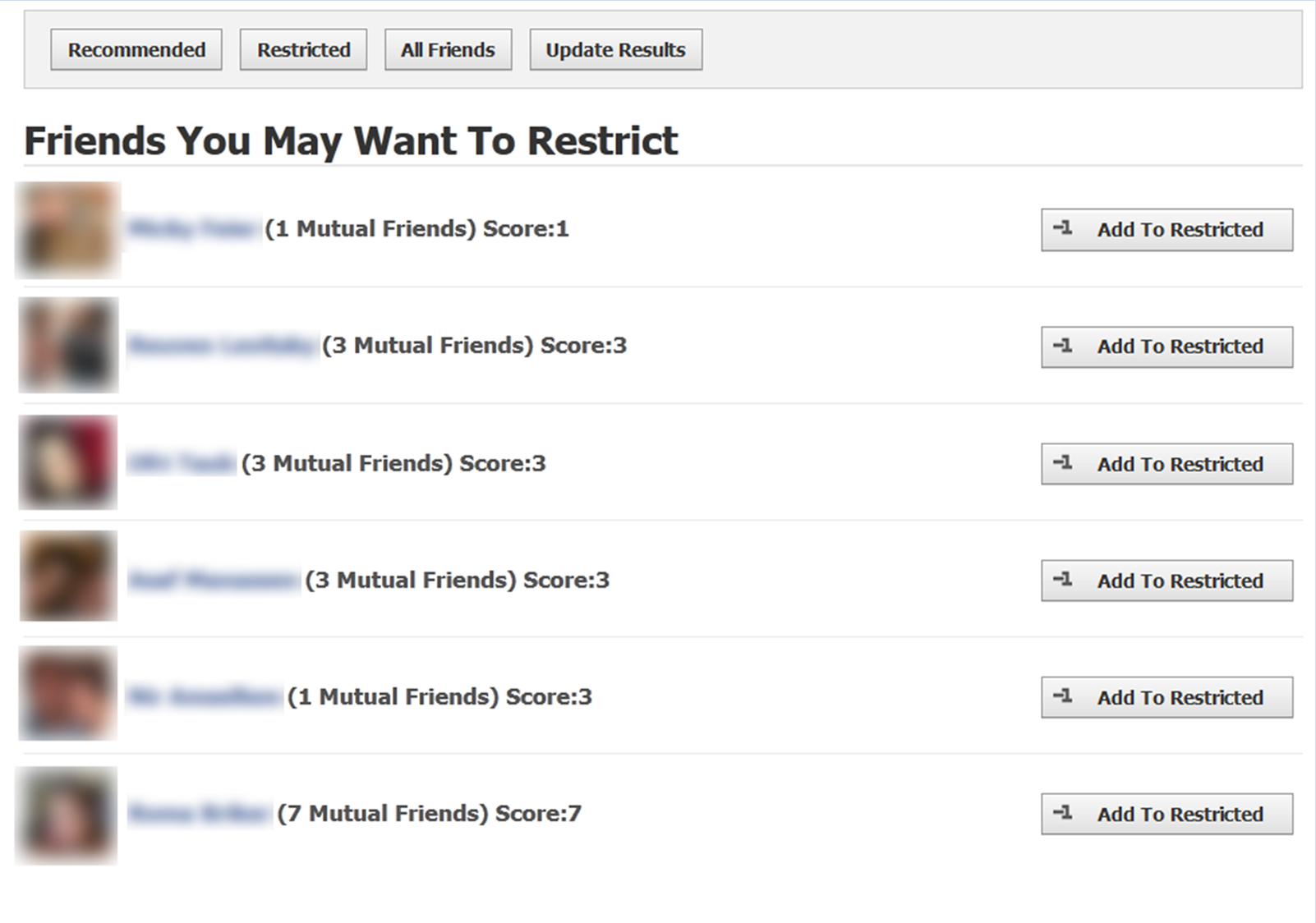}
\end{center}
\caption{Friends Analyzer Facebook application - user rank and sorted friends list.}
\label{faf}
\end{figure}

The Friends Analyzer Facebook application (also referred to as SPP application) is the part of the SPP responsible for analyzing a user friends list  to determine which of the user's friends may pose a threat to the user's privacy. 
After the user installs the Friends Analyzer application, the application scans the user's friends list and returns a credibility score for each one of the user's friends. Each friend's score is created by using simple heuristics or with a more sophisticated machine learning algorithm, which takes into account the strength of the connection between the user and his friends. The strength of each connection is based on different connection  features, such as  the number of common friends between the user and his friend and the number of pictures and videos the user and his friend were tagged in together (see Section~\ref{sec:methods}). At the end of the process, the user receives a web page, which includes a sorted list of all his friends according to each friend's score, where the friends with the lowest scores have the highest likelihood of being fake profiles appear on the top of the list (see Figure~\ref{faf}). 
For each friend in the returned sorted list, the user has the ability to restrict the friend's access to the user's private information by simply clicking on the restrict button attached to each friend in the sorted list. 
Moreover, the application provides the user an interface to view all his friends alphabetically and easily restricts access with a single click.
This option enables Facebook users to protect their privacy not only from fake profiles but also from real profiles, such as exes, for whom they do not want to have access to their personal data stored in their Facebook profile.

\subsection{Social Privacy Protector Firefox Add-on}
\begin{figure}[ht]
\begin{center}

\includegraphics[width=0.8\textwidth,clip]{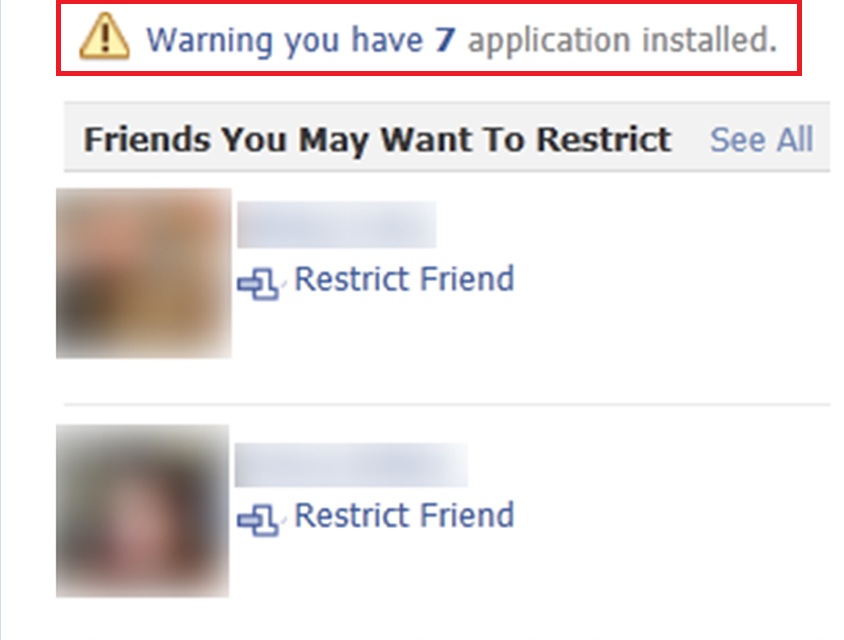}
\end{center}
\caption{Social Privacy Protector Firefox Add-on - warning about installed applications and friends you may want to restrict.
}
\label{addon2}
\end{figure}

The Social Privacy Protector Firefox Add-on (also referred to as Add-on) is the part of the SPP software responsible for improving user privacy settings with just a few simple clicks. After the Add-on is installed on the user's Firefox browser, it begins to monitor the user's internet activity. 
When the Add-on identifies that the user logged onto his Facebook account, the Add-on then analyzes the number of applications installed on the user's Facebook profile and presents a warning with the number of installed applications, which may pose a threat to the user's privacy (see Figure~\ref{addon2}). 
The Add-on also presents the top two results obtained by the Friends Analyzer Facebook application and suggests  which friends to restrict (see Figure~\ref{addon2}).

\begin{figure}[ht]
\begin{center}
\includegraphics[
width=1\textwidth,clip]{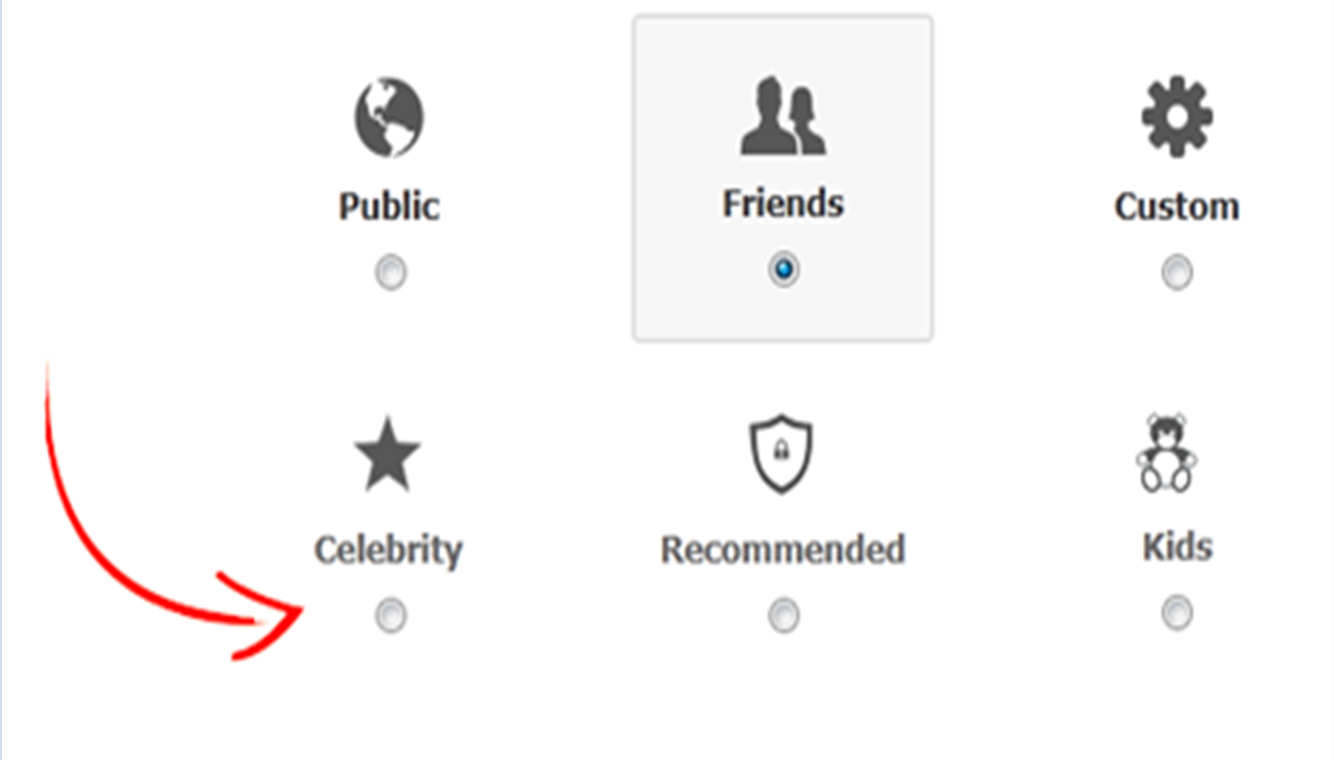}
\end{center}
\caption{Social Privacy Protector Firefox Add-on - optimizing the user's privacy setting with one simple click.}
\label{addon}
\end{figure}

The Add-on also detects when the user has entered Facebook's privacy settings page and presents the user with three new privacy setting options. The new privacy settings are based on the user's profile type and can be modified with one click (see Figure~\ref{addon}), instead of the more complex Facebook custom privacy settings that may contain up to 170 options~\cite{anwar2012visualization}.
Using the new Add-on privacy settings a user can simply chose the profile type most suitable for him out of three options: a) \textit{Celebrity setting} -  in this setting all of the user's information is public, b) \textit{Recommended setting} - in this setting the user's privacy is only public to friends, however some of the user's details, such as profile name and pictures, are public, and c) \textit{Kids settings} - in this setting the profile is only open to the user's friends and only friends of friends can apply for friend requests. Using this Add-on a user can easily control and improve their privacy without contacting a security expert.
Moreover, parents can simply install this Add-on on their children's Facebook accounts in order to better protect their children's privacy on Facebook without needing to understand Facebook's different privacy setting options.   
Our Add-on is also easy for customizing privacy settings by adding more privacy option settings to different types of users. 
Furthermore, it is easy to customize our privacy settings by adding more optional privacy settings for different types of users.
In this study, we utilized users data collected by the Add-on  to further study the privacy settings of Facebook users.

\subsection{HTTP Server}
The HTTP server is the part of the SPP  responsible for connecting the SPP Firefox Add-on to the SPP Facebook application.
When a user installs the SPP software, the server analyzes the user's friends list and identifies which of the user's friends may pose a threat on his security. 
Also, to enhance the application's performance, the HTTP server caches parts of the analyzed results.  In order to protect the user's privacy, the application stores only the minimal number of features in an encrypted manner using RC4 encryption.

\section{Methods and Experiments}
\label{sec:methods}
In this study our experiments are divided into two main parts. 
In the first part, we deployed an initial version of the SPP software in order to improve user privacy on Facebook. We also used the initial version to collect data on each SPP's user and his links. The main focus of this part is calculating the heuristic, which sorts the friends list and recommends which friends to restrict. Additionally, in this part, we also present methods for analyzing  the user privacy setting data collected by the Add-on to better understand Facebook user privacy settings and understand how much they are exposed to different security threats, such as information exposure to fake friends.

In the second part, we use the data collected in the first part to learn more about Facebook user's privacy settings. Furthermore, we used the collected data in order to develop machine learning classifiers, which can identify Facebook profiles with higher likelihoods of being fake.  Moreover, these classifiers can be used to replace the initial SPP heuristic with a more generic model, which can provide SPP users recommendations on which friends to restrict.  In the second part our main focus is on constructing the machine learning classifiers and evaluating their performances.

In the remainder of this section, we present in detail the methods used in each one of the two parts.

\subsection{Deploying Social Privacy Protector - Initial Version}
\label{spp_features}
After we developed the SPP's software according to the architecture described in Section~\ref{architecture}, we had to develop a heuristic, which can quickly sort the friends list of each SPP user. 
In the initial version, we developed the heuristic to be as simple as possible and based it upon the hypothesis that most fake users do not have strong connections with real users. To estimate the strength of a connection between two users, we extracted lists of features and calculated a simple arithmetic heuristic. The heuristic's main constrain was that for every SPP user, it needed to analyze and evaluate hundreds or even thousands of connection strengths between the user and each one of his friends in a short period of time. Moreover, the heuristic needed to take into consideration the performance of the Facebook application API~\cite{facebook-api}, in its calculation time of each feature. 
After testing and evaluating several different features, we decided to extract the following features for each SPP user (referred as $u$), and each one of his friends (referred as $v$):
\begin{enumerate}
\item \textbf{Are-Family(u,v)} - mark if $u$ and $v$ are defined in Facebook as being in the same family. The \textit{Are-Family} feature prevents the cases in which the SPP application will mark a family member as a fake profile.
\item \textbf{Common-Chat-Messages(u,v)} - the number of chat messages sent between $u$ and $v$. We assume that in most cases, such as in a fake profile used to send spam messages, there will be no chat interaction between the user and the fake profile. However, when there are different types of fake profiles, such as fake profiles used by cyber predators, this feature will be less helpful in identifying the threat.
\item \textbf{Common-Friends(u,v)} - the number of mutual friends both $u$ and $v$ poses. The relevance of the \textit{Common-friends} feature is very intuitive. It is expected that the larger the size of the common neighborhood, the higher the chances are that the friendship between the users is  real. The Common-Friends feature was previously used to solve different versions of the link prediction problem~\cite{liben,kahanda2009using,leskovec2010predicting,fire2011link} and was found to be a very useful feature in many of these scenarios.
\item \textbf{Common-Groups-Number(u,v)} - the number of Facebook groups both $u$ and $v$ are members in. It is expected that the higher the number of groups both users are member of, the higher the chances that $u$ and $v$ have similar fields of interest, which might indicate that the friendship between $u$ and $v$ is real. 
The Common-Groups-Number feature was used in the study of Kahanda and Neville~\cite{kahanda2009using} to predict link strength.
\item \textbf{Common-Posts-Number(u,v)} - the number of posts both $u$ and $v$ posted on each other's wall in the last year. Similar post features were studied by Kahanda and Neville and were discovered to be very useful in predicting strong relationships between two Facebook users~\cite{kahanda2009using}.
\item \textbf{Tagged-Photos-Number(u,v)} - the number of photos both $u$ and $v$ were tagged in together. We assume that most fake profiles have almost no shared tagged photos with the user. The Tagged-Photos-Number feature was also used in the Kahanda and Neville study~\cite{kahanda2009using}.
\item \textbf{Tagged-Videos-Number(u,v)} - the number of video clips both $u$ and $v$ appeared in together. As in the case of tagged photos, we assume that most fake profiles have almost no shared tagged video clips with the user.

\item \textbf{Friends-Number(u) and Friends-Number(v)} - the total number of friends $u$ and $v$ have. These features are usually referred to as the \textit{degree} of $u$ and the \textit{degree} of $v$, and were extracted from each user to assist us in improving the supervised learning classifiers, as described in Section~\ref{sec:ml}. However, we did not use this feature when calculating the connection strength heuristics in SPP's initial version.
\end{enumerate}

In an attempt to build better fake profile prediction heuristics, we also tested the following features: a) the number of mutual Facebook likes both $u$ and $v$ gave to each other, and b) the number of comments both $u$ and $v$ posted on each other's wall posts. However, although these two features seemed  promising in assisting us in identifying fake profiles, we did not use them in the end due to performance issues, i.e.,  calculating these two features was too time consuming and inappropriate to use in order to receive real-time results.

After several tests and simple evaluations, we decided to use the following simple heuristic in order to define the \textit{Connection-Strength} function between a user $u$ and its friend $v$:
\begin{eqnarray}
Connection\mbox{-}Strength(u,v) =CS(u,v) &:=&  Common\mbox{-}Friends(u,v) \nonumber  \\
&+&   Common\mbox{-}Chat\mbox{-}Messages(u,v) \nonumber \\
&+&   2\cdot Common\mbox{-}Groups\mbox{-}Number(u,v) \nonumber \\
&+&   2\cdot Common\mbox{-}Posts\mbox{-}Number(u,v)\nonumber \\
&+&   2\cdot Tagged\mbox{-}Photos\mbox{-}Number(u,v) \nonumber \\
&+&   2\cdot Tagged\mbox{-}Videos\mbox{-}Number(u,v) \nonumber \\
&+&   1000 \cdot Are\mbox{-}Family(u,v) \nonumber
\end{eqnarray}
In order to tell SPP users which friends ought to be restricted, we ranked each user's friends list according to the \textit{Connection-Strength(u,v)} function. Due to Facebook's estimations that 8.7\% of all Facebook users do not belong to real profiles~\cite{facebook_fake}, we presented to each SPP user the top 10\% of his friends who received the lowest Connection-Strength score (see illustration in Figure~\ref{faf}). 

To evaluate the performance of the Connection-Strength heuristic, we calculated three statistics on the heuristic's performances in restricting friends. First, we calculated the heuristic's restricting precision for different Connection-Strength values. 
This calculation was performed by measuring, for different Connection-Strength values, the ratio between the number of friends who were restricted and the total number of friends  who received the exact Connection-Strength value. Second, 
we calculated the restriction rates according to the friends' ranking positions in the restriction interface.
This calculation was performed by measuring the percentage of friends, which were restricted in each position in the restriction interface. 
Lastly, we also calculated the heuristic's average users precision at $k$ for different $k$ values, 
in the following manner.  First, for each SPP user $u$, which had at least $k$ friends, we calculated the user's Connection-Strength average precision at $k$. 
This was done by selecting $k$ friends, which received the lowest Connection-Strength values with $u$ out of all $u$'s friends\footnote{In case more than $k$ friends received the lowest Connection-Strength values, we randomly removed friends with the highest Connection-Strength values, until we were left with exactly $k$ friends.}. 
We then calculated the ratio between the number of friends $u$ had restricted among the selected friends and $k$. 
After we finished calculating the Connection-Strength average precision at $k$ for each user $u$, we then continued to 
calculate heuristic's average users precision at $k$, by simply summing up all the users' Connection-Strength average precision at $k$, and dividing the sum by the number of users with at least $k$ friends.
A formal arithmetical definition of the heuristic's average users precision at $k$ is as follows:
\[ CS\mbox{-}Avg\mbox{-}Precision(k) : = \frac{\Sigma_{\{u \in Users | |friends(u)| \geq k \}} P_{u}(k)}{|\{u \in Users | |friends(u)| \geq k \}|} \]
Where $Users$ is a set,  which contains all SPP users, $friends(u)$ is a set which contains all of $u$'s Facebook friends, and $P_{u}(k)$ defined to be the heuristic's precision at $k$ for a user $u$:
\[P_{u}(k) := \frac{\Sigma_{\{f \in friends(u) | \exists f_i,...,f_{n-k} \in friends(u),\forall j \in [i,..,n-k]  CS(f) \geq CS(f_j)  \}} is\mbox{-}restricted(u,f)} {k}.\]
Where $is\mbox{-}restricted(u,f)$ is a function, which returns 1 if a user $u$ had restricted his friend $f$ or 0 otherwise.

The goal of the presented Connection-Strength heuristic was not to identify fake profiles in an optimal way, rather to make Facebook users more aware of the existence of fake profiles and the fact that these types of profiles can threaten their privacy. Additionally, we wanted to collect a unique dataset that would contain labeled profiles with high likelihoods of being fake profiles. 

In addition to collecting meta-content features through the SPP application, the SPP Firefox Add-on also collected the following users defined privacy settings each time the user used the Add-on:
\begin{enumerate}
 \item \textbf{Installed-Application-Number} - the number of installed Facebook applications on the user's Facebook account.
  \item \textbf{Default-Privacy-Settings} - the default user's privacy settings on Facebook, which can be one of these values: public, friends or custom. 
  This setting is responsible for the scope in which content created by the user will be exposed by default. For example, if the default privacy settings are set to ``friends'' by default, only the user's friends can see his posted content. The SPP's Add-on only stored this value if the user's privacy was set to public or friends.
This privacy setting default value for new Facebook users is set to public.
  
   \item \textbf{Lookup} - regulates who can look up the user's Facebook profile by name. This setting can take one of the following values: everyone, friends or friends of friends.    
   This privacy setting default value for new Facebook users is set to everyone.
   \item \textbf{Share-Address} - this value is responsible for defining who can see the user's address. This setting can take one of the following values: everyone, friends or friends of friends. 
This privacy setting default value for new Facebook users is set to everyone.
   \item \textbf{Send-Messages} - this value is responsible for defining who can send messages to the user. This setting can take one of the following values: everyone, friends or friends of friends. 
   This privacy setting default value for new Facebook users is set to everyone.
   \item \textbf{Receive-Friend-Requests} - this value is responsible for defining who can send friend request to the user. This setting is limited to two values only : everyone and friends of friends. The default value for this setting is everyone.
	\item \textbf{Tag-Suggestions} - this value is responsible for defining which Facebook users will receive photo tag suggestions when photos that look like the user have been uploaded onto Facebook. This setting can take one of the following values: no one or friends. 
	This privacy setting default value for new Facebook users is set to friends.
   \item \textbf{View-Birthday} - this value is responsible for defining who can view the user's birthday. This setting can take one of the following values: everyone, friends or friends of friends. 
   This privacy setting default value for new Facebook users is set to friends-of-friends.

\end{enumerate}
By analyzing and monitoring the privacy settings, we can learn more about the SPP user's privacy settings on Facebook. In addition, we can estimate how vulnerable Facebook users' information is to fake profile attacks. 
Furthermore, by analyzing the collected privacy settings, we can also identify other potential privacy risks, which are common to many different users.  

\subsection{Supervised Learning}
\label{sec:ml}
After, we deployed the SPP software and gathered enough data on which friends SPP users had restricted, and which friends they had not restricted, our next step was to use supervised learning techniques  to construct fake profile identification classifiers.
To construct the fake profile identification classifiers, we first needed to define the different datasets and their underlining features. Next, we used different supervised learning techniques to construct the classifiers. Lastly, we evaluated the classifiers using different evaluation methods and metrics. 

In the remainder of this section we describe, in detail, the process of constructing and evaluating our classifiers. First, in Section~\ref{sec:datasets}, we describe how we defined the different datasets and their features. Second, in Section~\ref{sec:construct}, we 
describe which methods were used  to construct our classifiers and evaluate their performance. 
\subsubsection{Datasets and features}
\label{sec:datasets}
The SPP application's initial version collected and calculated many different details about each connection between a SPP user and each one of his friends in real-time (see Section~\ref{spp_features}). 
Moreover, the SPP application presented the user with two interfaces for restricting his friends. The first restriction  interface (referred to as the recommendation interface) presents the user with a list with the 10\% of his friends who received the lowest Connection-Strength score. The second  restriction interface (referred to as the alphabetical interface) presents  the user with all of his friends in alphabetical order. Using these two restriction interfaces, we defined four types of links sets, two unrestricted links sets, and two restricted links sets:
\begin{enumerate}
\item \textbf{All unrestricted links set} - this set consists of all the links between the application users and their Facebook friends
who were not restricted by the application.
\item \textbf{Recommended unrestricted links set} - this set consists of all the links between the application users and their Facebook friends who were recommended for restriction by the application due to a low Connection-Strength score, but who were not restricted by the user.
\item \textbf{Recommended restricted links set} - this set consists of all the links between the application users and their Facebook friends who were recommended for restriction by the application due to a low Connection-Strength score and who were restricted by the user.

\item \textbf{Alphabetically restricted links set} - this set consists of all the links between the application users and their Facebook friends who were not recommended for restriction by the application. However, the user deliberately chose to restrict them by using the alphabetical interface\footnote{If a restricted user's friend was presented in the recommendation interface and was restricted by using the alphabetical interface, the link between the user and the restricted friend was assigned to the recommended restricted links set.}.
\end{enumerate}

Using the defined above links sets, we define the following three  datasets:
\begin{enumerate}

\item \textbf{Fake profiles dataset} -  this dataset contains all the links in the \textit{Recommended unrestricted links set} and in the \textit{All unrestricted links set}. Namely, this dataset contains all friends who were restricted due to a relatively low Connection-Strength and all friends who were not restricted. Therefore, we believe
that this dataset is suitable  for constructing classifiers,  which can predict friends, who mostly represent fake profiles, the user  
need to restrict. We believe that this dataset is suitable for replacing the Connection-Strength heuristics with a generic classifier, 
which can recommend to a SPP user, which friends to restrict. In addition, the classifiers constructed from this type of dataset can assist online network administrators in identifying fake profiles across the entire the network.

\item \textbf{Friends restriction dataset} - this dataset contains all the links in the \textit{alphabetically restricted links set}, and in the
\textit{All unrestricted links set}. Namely, this dataset contains
all the friends who were not restricted and all the friends who were restricted deliberately by the user, although they were not recommended by the SPP application. 
Therefore, we believe
that this dataset is suitable  for constructing classifiers, which can predict friends, who mostly represent real profiles the user  
prefers to restrict. 

\item \textbf{All links dataset} - this dataset contains all the links in all four disjoint links sets. 
According to the dataset definition, this dataset is the largest among all defined datasets, we believe that like the Fake profiles dataset this dataset can be suitable for replacing the Connection-Strength heuristics with a generic classifier, which can recommend to a SPP user which friends to restrict.

\end{enumerate}

For each link in the above defined links datasets, the SPP application calculated all of the 8 first link features defined in Section~\ref{spp_features} in real-time including the Friends-Number($v$)\footnote{In some cases we were not able to extract the user's ($v$) friends number probably due to the $v$'s  privacy settings.} (referred to as the \textit{Friend Friends-Number}). In addition, if it was arithmetically possible, we also calculated the following set of seven features:
\begin{enumerate}
\item \textbf{Chat-Messages-Ratio(u,v)} - the ratio between the number of chat message $u$ and $v$ sent to each other and the the total number of chat messages $u$  sent to all of his friends.
The formal \textit{Chat-Messages-Ratio} definition is:
\[ Chat\mbox{-}Messages\mbox{-}Ratio(u,v) := \frac{Common\mbox{-}Chat\mbox{-}Messages(u,v)}{\sum_{f \in friends(u)}Common\mbox{-}Chat\mbox{-}Messages(u,f)}\]
Where $friends(u)$ is defined to be  a set which contains all the friends of $u$.

\item \textbf{Common-Groups-ratio(u,v)} - the ratio between the number of Facebook groups both $u$ and $v$ have in common and the maximum number of groups which $u$ and all of his friends have in common.
The formal \textit{Common-Groups-Ratio} is:
\begin{eqnarray}
 Common\mbox{-}Groups&\mbox{-}&Ratio(u,v) := \nonumber \\ 
 && \frac{Common\mbox{-}Groups\mbox{-}Number(u,v)}{max(\{Common\mbox{-}Groups\mbox{-}Number(u,f)|f \in friends(u)\})} \nonumber 
\end{eqnarray}
\item \textbf{Common-Posts-Ratio(u,v)} - the ratio between the number of posts both $u$ and $v$ posted on each others walls and the total number of posts which $u$ posted on all his friends' walls.
The formal \textit{Common-Posts-Ratio} is:
\[ Common\mbox{-}Posts\mbox{-}Ratio(u,v) := \frac{Common\mbox{-}Posts\mbox{-}Number(u,v)}{\sum_{f \in friends(u)}Common\mbox{-}Posts\mbox{-}Number(u,f)}\]

\item \textbf{Common-Photos-Ratio(u,v)} - the ratio between the number of tagged photos both $u$ and $v$ were tagged in together  and the total number of photos, which $u$ were tagged in.
The formal \textit{Common-Photos-Ratio} is:
\[ Common\mbox{-}Photos\mbox{-}Ratio(u,v) := \frac{Common\mbox{-}Photos\mbox{-}Number(u,v)}{\sum_{f \in friends(u)}Common\mbox{-}Photos\mbox{-}Number(u,f)}\]

\item \textbf{Common-Video-Ratio(u,v)} - the ratio between the number of videos both $u$ and $v$ were tagged in together on and the total number of videos, which $u$ were tagged in.
The formal \textit{Common-Video-Ratio} is:
\[ Common\mbox{-}Video\mbox{-}Ratio(u,v) := \frac{Common\mbox{-}Video\mbox{-}Number(u,v)}{\sum_{f \in friends(u)}Common\mbox{-}Video\mbox{-}Number(u,f)}\]

\item \textbf{Is-Friend-Profile-Private(v)} - in some cases the SPP application did not succeed in collecting \textit{v}'s friends number (\textit{Friends-Number(v)}),  and succeeded in collecting the \textit{Common-Friends(u,v)} value, which returned a value greater than zero. This case may indicate that \textit{v}'s profile is set to be a private profile. 
With these cases in mind, we defined the \textit{Is-Friend-Profile-Private} function to be a binary function, which returns true values in case the application did not succeed in collecting \textit{v}'s friends number and succeeded in collecting the \textit{Common-Friends(u,v)} with a value greater than zero, or a false value otherwise.

\item \textbf{Jaccard's-Coefficient(u,v)} - Jaccard's-coefficient is a well-known feature for link prediction \cite{liben,kahanda2009using,fire2011link}. The Jaccard's coefficient is defined as the number of common-friends $u$ and $v$ have divided by the sum of distinct friends both $u$ and $v$ have together. 
The formal \textit{Jaccard's-Coefficient} definition is:
\begin{eqnarray}
	Jaccard's&\mbox{-}&Coefficient(u,v):= \nonumber \\
	 && \frac{Common\mbox{-}Friends(u,v)}{Friends\mbox{-}Number(u)+Friends\mbox{-}Number(v) - Common\mbox{-}Friends(u,v)} \nonumber
\end{eqnarray}
 A higher value of \textit{Jaccard's-coefficient} denotes a stronger link between two Facebook users.

\end{enumerate}

\subsubsection{Classifiers Construction and Evaluation}
\label{sec:construct}
Using the three datasets and the 15 features defined in the previous sections, we constructed classifiers for fake profile identification and for recommending profiles for restriction. 

The process of constructing and evaluating the different classifiers was as follows. 
First we matched the suitable datasets for each type of classification mission in the following manner: 
a) for identifying fake profiles we used \textit{Fake profiles dataset},
b) for recommending real profiles for restriction we used the \textit{Friends restriction dataset}, 
and c) for replacing recommending to SPP users which real and fake friends to restrict we used the \textit{All links dataset}.
Next, for each link in each one of the datasets, we extracted the 15 features defined in Sections~\ref{spp_features} and ~\ref{sec:datasets}, and created a vector set for each link. 
Furthermore, we added  an additional binary target feature that indicates if the link between the SPP user and his friend was restricted by the SPP user to each link's features vector set.
Due to the fact that the majority of the links in each dataset had not been restricted, these datasets were overwhelmingly imbalanced with restricted links as a minority class. 
Therefore, a naive algorithm that always predicts ``not restricted'', will present good prediction precision.
To overcome the datasets imbalance problem we used a similar undersampling methodology  used by Guha et al. to predict trust~\cite{guha2004propagation} and by  Leskovec et al.~\cite{leskovec2010predicting} to predict positive and negative links. 
According, to this methodology we transform each imbalanced dataset into a balanced dataset by combing all the restricted links in each dataset and adding to them an equal number of randomly selected  unrestricted links from each dataset.
Afterwards, we used the balanced datasets with the updated extracted links' features vector sets to construct several classifiers by 
using WEKA~\cite{weka}, a popular suite of machine learning software written in Java and developed at the University of Waikato, New Zealand. 
We used WEKA's  C4.5 (J48), IBk, NaiveBayes, Bagging, AdaBoostM1, RotationForest, and RandomForest implementations of the corresponding algorithms.  In addition, we used the simple OneR classifier as a baseline for the performance of the other classifiers.
For each of these algorithms, most of the configurable parameters were set to their default values
with the following exceptions: for C4.5, the minimum number of instances
per leaf parameter was between the values of 2, 6, 8 and 10; for IBk, its k
parameter was set to 5 and 10; 
The ensemble methods were configured as follows: The number of iterations for all ensemble methods was set to 100. The Bagging, AdaBoostM1, and RotationForest algorithms were evaluated using J48 as the
base classifier with the number of instances per leaf set to 4, 6, 8, and 10.
Next, we evaluated our classifiers using the common 10-folds cross validation approach.
We used the area-under-curve (AUC), f-measure, true-positive and false-positive rate to evaluate the different classifiers' performances.
Additionally, in order to obtain an indication of the usefulness of the various features, we also analyzed the features 
importance by using WEKA’s information gain attribute selection algorithm.

Furthermore, in order to evaluate the classifiers recommendations precision at top \textit{k} (\textit{precision@k}), we selected the machine learning algorithm, which presented the highest AUC in the above evaluations and used two evaluation methods to measure the performance of the algorithm on the different datasets. In the first evaluation method, we split our datasets into training sets and testing sets. For each one of the three balanced datasets, we randomly split each dataset into a training dataset, which contained $\frac{2}{3}$ of the labeled instances and a testing dataset, which contained $\frac{1}{3}$ of the labeled instances. Afterwards, we constructed classifiers using the training dataset only.  Next, we used the classifiers to classify the profiles in the testing dataset and sorted  the instances according to the classifiers' prediction probabilities in descending order, where the links, which received the highest probability of being restricted were first. We then evaluated the classifiers' predictions precisions for the top \textit{k} predictions, for different values of \textit{k}.

In the second evaluation method, our goal was to measure the classifiers recommendations average users precision at \textit{k}. 
To achieve this goal we used the following method. First, we selected a user out of all SPP users. Afterwards, we created a training dataset using all SPP users' links without the selected user's links. Next, we balanced the training dataset using the same undersampling method described above. Afterwards, we constructed a classifier using the training dataset and used the selected user's links as a testing dataset. We used the constructed classifier to predict the probability of restriction for each link in the selected user's links.
We then sorted the classifier's predictions in descending order where the links, which received the highest probability of being restricted  were first. Subsequently, we measured the classifier's predictions precision for different \textit{k} values. 
Lastly, we repeated this process for each one of all SPP users and calculated the average classifiers' precisions for different
 \textit{k} values. 

Using these evaluation methods, we were able evaluate how precise our classifiers are in recommending which friends to restrict both from the SPP users' point of view, and from the online social network administrator point of view.

\section{Results}
\label{results}
The initial version of the SPP software was formally launched at the end of June 2012 as free to use software~\cite{spp,fire2012social}. The software launch received massive media coverage with hundreds of online articles and interviews in leading blogs and news websites, such as Fox news~\cite{foxnews} and NBC news~\cite{nbcnews}. 

Due to the media coverage, from the 27th of June, 2012 to the 10th of November, 2012, 3,017 users from more than twenty countries installed the SPP application out of which 527 users used the SPP application to restrict 9,005 friends, with at least one friend restricted for each user. In addition, more than 1,676 users had installed the SPP Firefox Add-on and removed at least 1,792 applications. 

In the remainder of this section we present the results obtained from analyzing the collected SPP software data in the following manner.
First, in Section~\ref{sec:results_app}, we present the datasets obtained by the SPP application. Afterwards, in Section~\ref{sec:results_ml}, we present the results obtained by our machine learning classifiers. 
Lastly, in Section~\ref{sec:results_addon} we present statistics on the success of our Add-on to assist Facebook users in removing unneeded applications from their profiles. Furthermore, in this section, we also presented different statistics about Facebook user privacy settings obtained from the SPP Add-on.

\subsection{Collected Datasets}
\label{sec:results_app}
After the initial software launch the SPP application was installed by 3,017 users out of which 527 users had restricted 9,005 friends. All friends were restricted between the 27th of June, 2012 and the 10th of November, 2012. 
To our great surprise 355 SPP application users used the application not only to remove users that received low Connection-Strength score, but to also search and restrict specific friends that are probably real profiles by name.  

Using the collected users' data we created three datasets as described in Section~\ref{sec:datasets} (see Table~\ref{table:datasets}). 
The first dataset was the Fake-profiles dataset, this dataset contained 141,146 
out of which the 434 SPP users had restricted 2,860 links (2.03\% of all links), which were recommended by the SPP application.
The second dataset was the Friends-restriction dataset, this dataset contained 144,431 links out of which the  355 users had restricted 6,145 links (4.25\% of all links), that were specifically chosen for restriction by the users using the Alphabetical-interface. The last  dataset was the All links dataset, which contained 151,825 links out of which 9,005 (6.01\% of all links), were restricted.  As expected all three datasets were overwhelmingly imbalanced with imbalance rates ranging from 2.03\% to 6.01\%. 

\begin{table}[ht]
\caption{Links Datasets\label{table:datasets}}{
\begin{tabular}{ |c |c| c| c| c|  } 
\hline 
	 	   &\textbf{Users Number} & \textbf{Restricted Link}  &\textbf{Unrestricted Links} & \textbf{Total Links}\\  

\hline 
\textbf{Fake-Profiles}  &434& 2,860 (2.03\%) & 138,286 & 141,146\\

\hline
\textbf{Friends Restriction} &355  & 6,145 (4.25\%) & 138,286 & 144,431\\
\hline
\textbf{All Links} &527&  9,005 (6.01\%) & 138,286 & 147,291\\
\hline

\end{tabular}
}
\end{table}

To better understand the differences between the restricted links features and  unrestricted links features, we calculated the average values of each extracted feature in each dataset for each link type (see Table~\ref{table:avg_features}). It can be noted that in all
the examined features, except for the \textit{Friend Friends-Number} and \textit{Is-Friend-Profile-Privat}e features, the restricted links features received a lower average than the unrestricted links features in each dataset.

\begin{figure}[ht]
\begin{center}

\includegraphics[
 width=0.8\textwidth,clip]{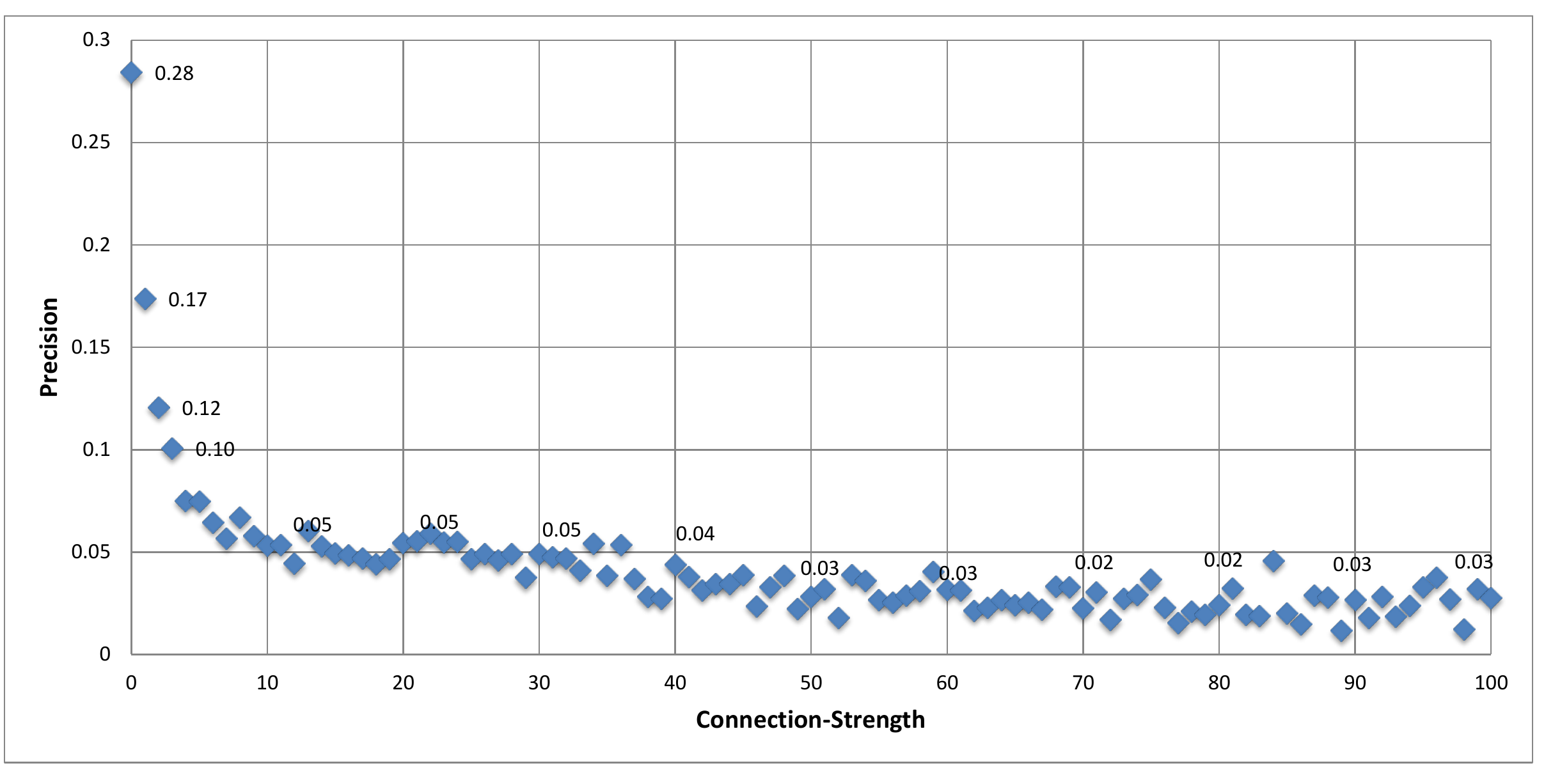}
\end{center}
\caption{ Friends restriction precision for different Connection-Strength values - it can be noted that among all users' friends, which received a Connection-Strength of 3, only 10\% were actually restricted.}
\label{fig:cs_stat}
\end{figure}

\begin{figure}[ht]
\begin{center}

\includegraphics[
 width=0.8\textwidth,clip]{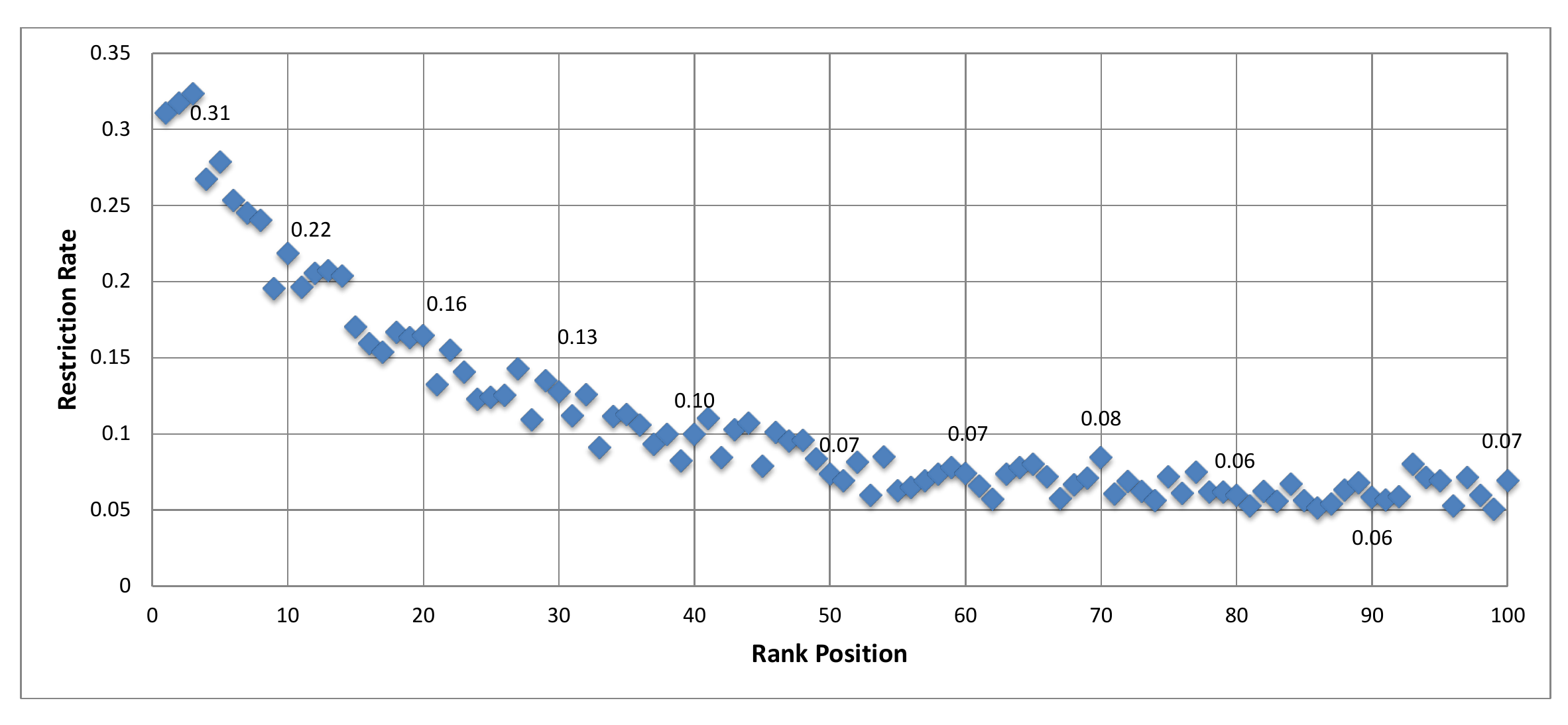}
\end{center}
\caption{ Connection-Strength restriction rates according to  friends' ranking positions in the Restriction interface - it can be noted that among all friends, which were ranked in the first position in the friends Restriction interface 31.1\% were actually restricted by the SPP users.}
\label{fig:ranking_rates}
\end{figure}

\begin{figure}[ht]
\begin{center}

\includegraphics[
 width=0.8\textwidth,clip]{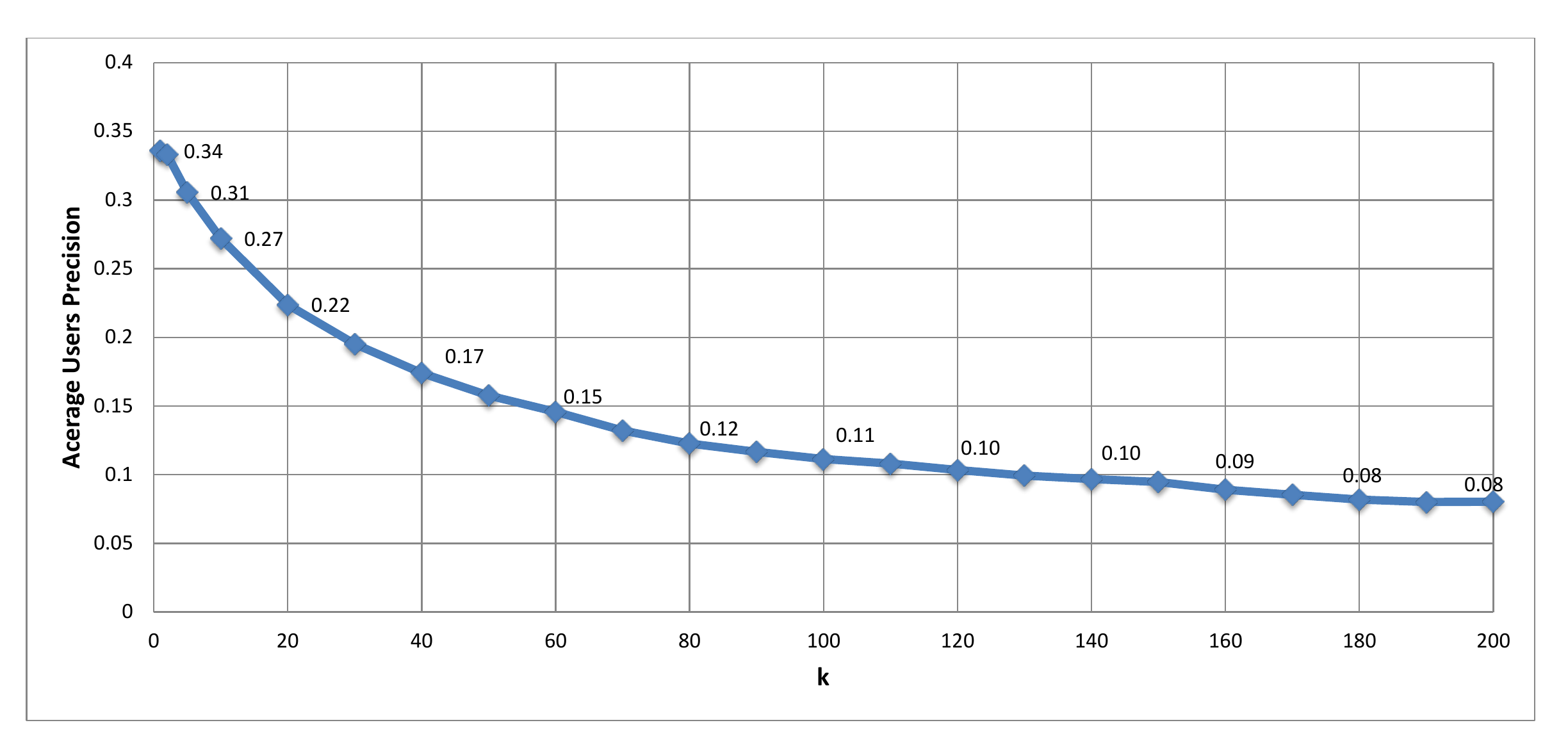}
\end{center}
\caption{ Connection-Strength average users precision at $k$ - it can be noted that the Heuristic's average users precision at 1 and  average users precision at 100 was 33.6\% and 11.1\% respectively.}
\label{fig:cs_avg}
\end{figure}

\begin{table*}[ht] 

\caption{  Features Average Values for  Different Datasets\label{table:avg_features}}{ 
\centering 
\begin{tabular}{|c | c |c |c| c |} 
\hline
\textbf{Feature}  & \textbf{Link Type}  &\textbf{Fake} & \textbf{Friends} & \textbf{All Links}    \\
    	 &  &\textbf{Profiles} & \textbf{Restriction}  &  \\
\hline
\multirow{2}{*}{\textbf{Are-Family}}& Restricted & 0  & 1 link &   1 link  \\ 
                           & Unrestricted & 9 links   &  9 links  & 9 links \\
\hline
\multirow{2}{*}{\textbf{Common-Chat-Messages}}& Restricted & 0.02 & 6.35 &  4.34 \\
                           & Unrestricted & 30.86 & 30.86 &30.86 \\
\hline
\multirow{2}{*}{\textbf{Common-Friends}}& Restricted & 1.44 &  19.8 & 13.97 \\ 
                           & Unrestricted & 36.78 & 36.78 &36.78 \\

\hline
\multirow{2}{*}{\textbf{Common-Groups-Number}}& Restricted & 0.028 & 0.56 & 0.392 \\
                           & Unrestricted & 0.689 & 0.689 &0.689 \\ 
\hline
\multirow{2}{*}{\textbf{Common-Posts-Number}}& Restricted & 0.008 &  0.069 & 0.049 \\
                           & Unrestricted & 0.147 &  0.147& 0.147\\
\hline
\multirow{2}{*}{\textbf{Tagged-Photos-Number}}& Restricted & 0.004 & 0.208 & 0.143 \\
                           & Unrestricted & 0.3 & 0.3& 0.3 \\
\hline
\multirow{2}{*}{\textbf{Tagged-Videos-Number}}& Restricted & 0 & 0.007& 0.005 \\
                           & Unrestricted & 0.017 & 0.017 & 0.017\\
\hline
\multirow{2}{*}{\textbf{Friend Friends-Number}}& Restricted & 627.31 & 819.57 & 756.25  \\
                           & Unrestricted & 703.31  & 703.31 & 703.31 \\
\hline
\multirow{2}{*}{\textbf{Chat-Message-Ratio}}& Restricted & $2.46 \cdot 10^{-5}$ &  0.003 & 0.002 \\
                                   & Unrestricted & 0.004  & 0.004 & 0.004 \\
\hline
\multirow{2}{*}{\textbf{Common-Groups-Ratio}}& Restricted & 0.006 &   0.108 & 0.076 \\
                           & Unrestricted & 0.118  & 0.118 & 0.118 \\
\hline
\multirow{2}{*}{\textbf{Common-Posts-Ratio}}& Restricted & 0.0003 & 0.003 & 0.002  \\
                           & Unrestricted & 0.0035  & 0.0035 & 0.0035\\

\hline
\multirow{2}{*}{\textbf{Common-Photos-Ratio}}& Restricted & $2.23 \cdot 10^{-5}$ & 0.003 & 0.002   \\
                           & Unrestricted &   0.0034& 0.0034 &0.0034 \\

\hline
\multirow{2}{*}{\textbf{Common-Video-Ratio}}& Restricted & 0 &  0.001 & 0.0007 \\
                           & Unrestricted &   0.0027& 0.0027 &0.0027 \\

\hline
\multirow{2}{*}{\textbf{Is-Friend-Profile-Private}}& Restricted & 5.87\% &  10.79\% & 9.23\% \\
                           & Unrestricted & 9.81\%  & 9.81\% & 9.81\%\\
\hline
\multirow{2}{*}{\textbf{Jaccard's-Coefficient}}& Restricted & 0.003 &  0.034 & 0.024 \\
                           & Unrestricted & 0.045  & 0.045 & 0.045\\
\hline

\end{tabular} 

}
\end{table*}
To understand how well the Connection-Strength heuristic performed, we calculated, as described in Section~\ref{spp_features}, the heuristic's restricting precision for different Connection-Strength values (see Figure~\ref{fig:cs_stat}), the heuristic restriction rates according to the friends' ranking positions in the Restriction interface (See Figure~\ref{fig:ranking_rates}), and the heuristic's average users precision at $k$  for different values of \textit{k} (see Figure~\ref{fig:cs_avg}).

Although the Connection-Strength heuristic was quite simple, it presented an average users precision of 33.6\% at 1, an average users precision of 27.1\% at 10, and an average users precision of 11\% at 100 (see Figure~\ref{fig:cs_avg}). In addition, 31.7\% of the friends, which  appeared in the second position in the Restriction interface, due to a low Connection-Strength score,  were actually restricted by the SPP users (See Figure~\ref{fig:ranking_rates}). Furthermore, 28\% of the SPP users' friends who received a Connection-Strength of 0 were also restricted (see Figure~\ref{fig:cs_stat}). However, the friends' restriction rates sharply declined when the Connection-Strength score increased. For example, only 10\%, of the users' friends, which received a Connection-Strength equal to 3 were actually restricted.

\subsection{Classifiers' Results}
\label{sec:results_ml}
\begin{figure}[ht]
\begin{center}

\includegraphics[
 width=0.8\textwidth,clip]{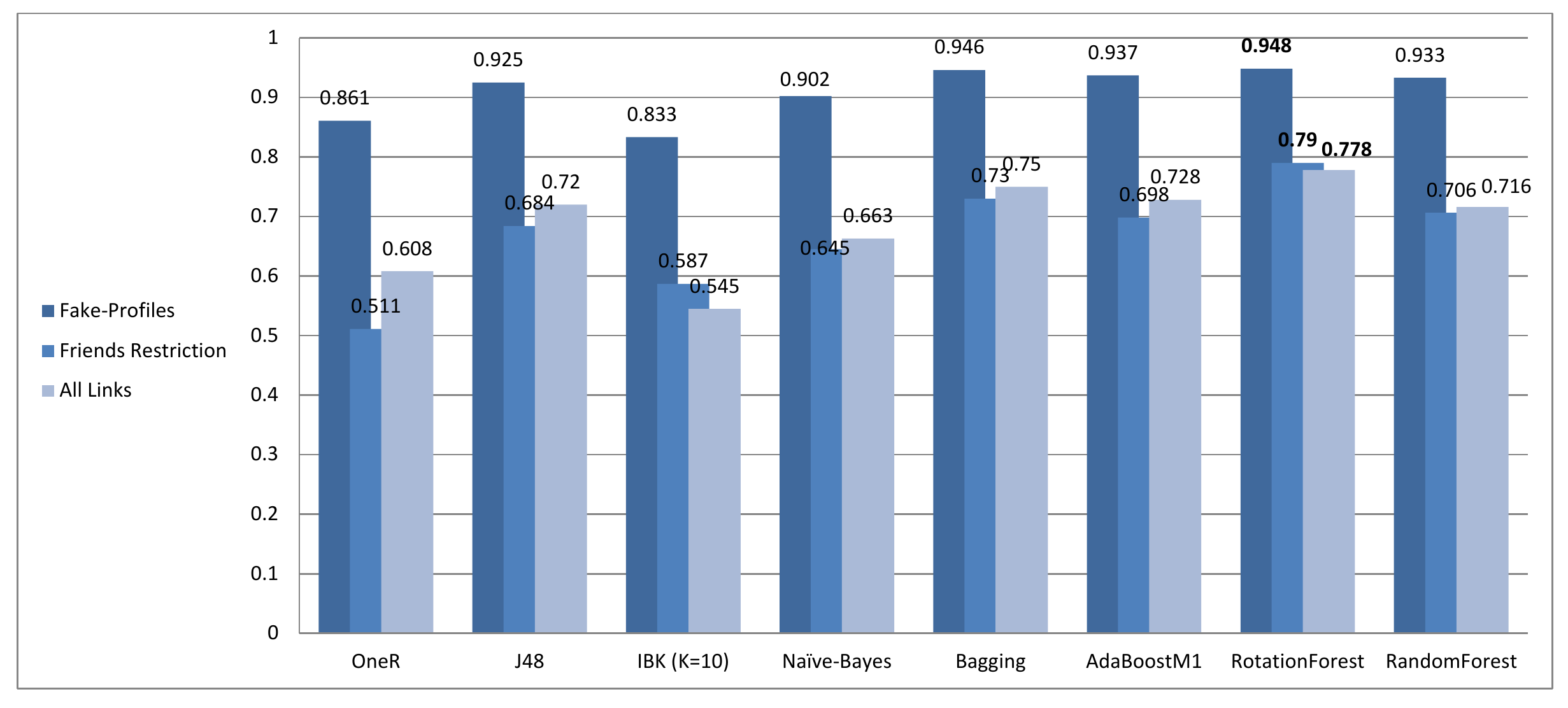}
\end{center}
\caption{ Classifiers AUC on the Different Datasets - it can be noted that the Rotation-Forest classifier received the highest AUC rates on all three datasets.}
\label{fig:auc}
\end{figure}

From the three imbalanced datasets, we created three balanced datasets by using all the restricted links in each dataset and randomly choosing an equal amount of unrestricted links. We then used the balanced dataset and  evaluated the specified machine learning
algorithms (see Section~\ref{sec:construct}) using a 10-fold cross-validation approach.  
The evaluation results of the different classifiers are presented in Figure~\ref{fig:auc} and in Table~\ref{table:10_folds}. 
It can be seen that on all datasets the Rotation-Forest classification algorithm presented the best AUC results among all the ensemble classifiers and the J48 decision tree classification algorithm presented the best results among all the non-ensemble classifiers. In addition, it can be noted that on all datasets the Rotation-Forest classifier presented considerably better results than the simple OneR classifier, which we used as baseline.

\begin{table*}[ht] 

\caption{ Classifiers' Performance on the Different Datasets  \label{table:10_folds}}{ 
\centering 
\begin{tabular}{|c | c |c |c| c|} 
\hline
\textbf{Classifier} & \textbf{Measure}  &\textbf{Fake Profiles} & \textbf{Friends Profiles} & \textbf{All Links}   \\
					&					& & \textbf{Restriction} &  \\

\hline
\multirow{4}{*}{\textbf{OneR}}& AUC 			 & 0.861   & 0.511 &  0.608 \\
                           	& F-Measure 	 & 0.867   & 0.531 &  0.616 \\
                           	& False-Positive & 0.179   & 0.532 &  0.414  \\
                           	& True-Positive &  0.902   & 0.554 &  0.623\\

\hline
\multirow{4}{*}{\textbf{J48}}& AUC 			 & 0.925   & 0.684 &  0.72 \\
                           	& F-Measure 	 & 0.885   & 0.668 &  0.659 \\
                           	& False-Positive & 0.179   & 0.498 &  0.321 \\
                           	& True-Positive &  0.937   & 0.754 &  0.645 \\
\hline                           	
\multirow{4}{*}{\textbf{IBK (K=10)}}& AUC 	 & 0.833   & 0.587 &  0.545 \\
                           	& F-Measure 	 & 0.744   & 0.49 &  0.637 \\
                           	& False-Positive & 0.174   & 0.289 &  0.749  \\
                           	& True-Positive &  0.696   & 0.419 &  0.817 \\
\hline  	
\multirow{4}{*}{\textbf{Naive-Bayes}}& AUC 	 & 0.902   & 0.645 &  0.663 \\
                           	& F-Measure 	 & 0.833   & 0.678 &  0.287 \\
                           	& False-Positive & 0.373   & 0.856 &  0.055  \\
                           	& True-Positive &  0.979   & 0.955 &  0.177 \\
\hline
\multirow{4}{*}{\textbf{Bagging}}& AUC 	 & 0.946   & 0.73 &  0.75 \\
                           	& F-Measure 	 & 0.89   & 0.677 &  0.675 \\
                           	& False-Positive & 0.171   & 0.403 &  0.3  \\
                           	& True-Positive &  0.938   & 0.717 &  0.662 \\
\hline
\multirow{4}{*}{\textbf{AdaBoostM1}}& AUC 	 & 0.937   & 0.698 &  0.728 \\
                           	& F-Measure 	 & 0.882   & 0.645 &  0.657 \\
                           	& False-Positive & 0.163   & 0.403 &  0.312  \\
                           	& True-Positive &  0.941   & 0.671 &  0.643 \\
\hline
\multirow{4}{*}{\textbf{Rotation-Forest}}& AUC 	 & 0.948   & 0.79 &  0.778 \\
                           	& F-Measure 	 & 0.897   & 0.719 &  0.696 \\
                           	& False-Positive & 0.158   & 0.336 &  0.275  \\
                           	& True-Positive &  0.941  & 0.75 &  0.681 \\
\hline
\multirow{4}{*}{\textbf{Random-Forest}}& AUC 	 & 0.933   & 0.706 &  0.716 \\
                           	& F-Measure 	 & 0.858   & 0.613 &  0.663 \\
                           	& False-Positive & 0.14   & 0.278 &  0.369 \\
                           	& True-Positive &  0.857  & 0.565 &  0.679 \\
\hline
\end{tabular} 

}
\end{table*}

\begin{figure}[ht]
\begin{center}

\includegraphics[
 width=0.8\textwidth,clip]{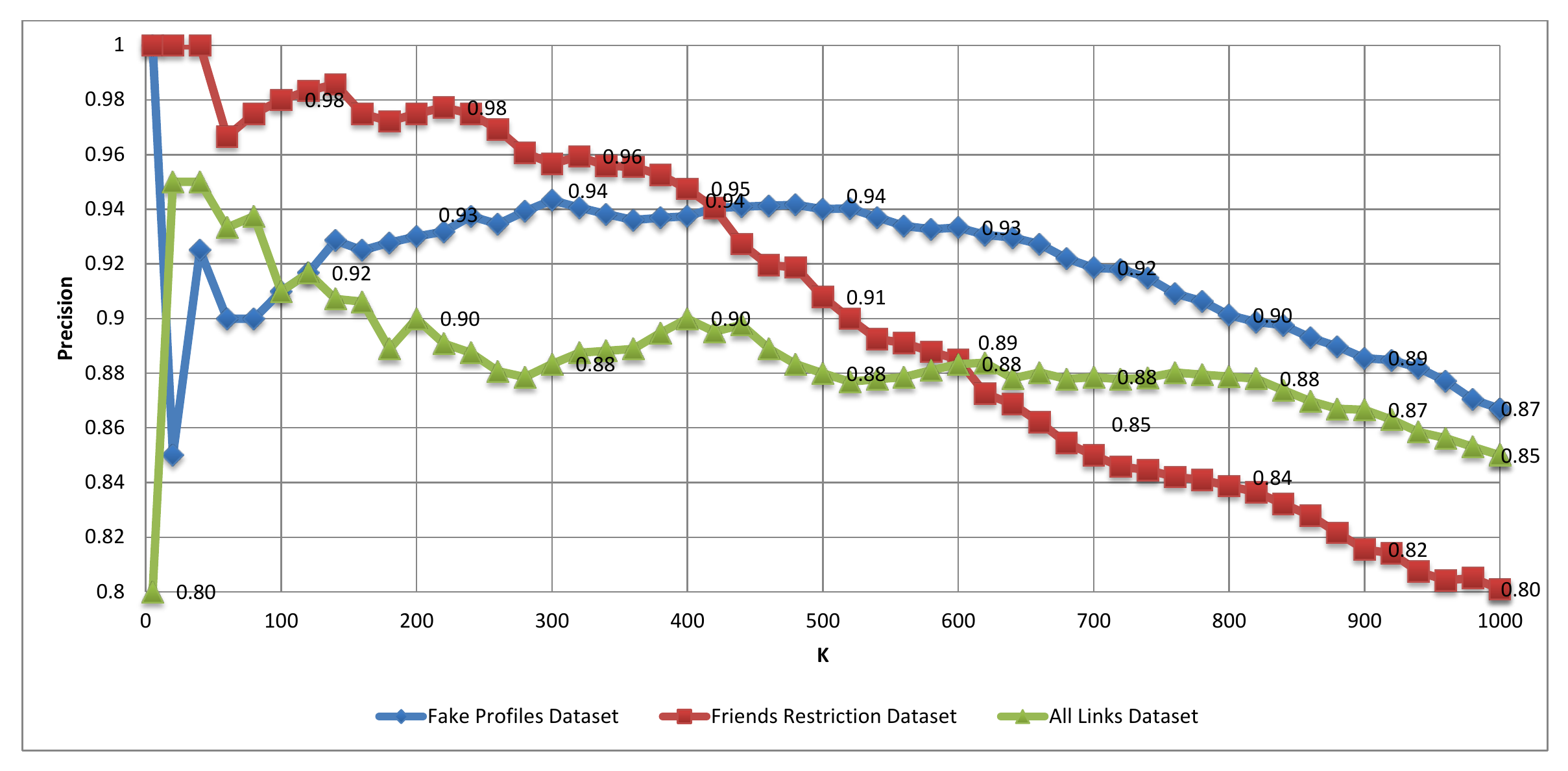}
\end{center}
\caption{Rotation-Forest Precision@k - it can been seen that the classifiers' precision at 100 was 98\%, 91\%, and 91\% for the Friends restriction dataset, Fake profiles datasets, and All links dataset respectively.}
\label{fig:precision_k}
\end{figure}

After, we discovered that the Rotation-Forest classifier presented the best overall results, we evaluated the Rotation-Forest classifier precision at \textit{k} for different values of \textit{k} on the different datasets. 
We first calculated  the classifier precision for different \textit{k} values by splitting each dataset into a training dataset, which contained $\frac{2}{3}$  of the links and a testing dataset, which contained $\frac{1}{3}$ of the links. 
The results of this precision at \textit{k} evaluation are presented in Figure~\ref{fig:precision_k}. 
It can be noted that the Rotation-Forest classifiers presented precision at 200 of  98\%, 93\%, and 90\% for the Friends restriction dataset, Fake profiles datasets, and All links dataset respectively. In addition, it can been noted that the classifiers' precision at 500 was 94\%, 91\%, and 88\% for the Fake profiles datasets, Friends restriction dataset, and All links dataset. Hence, out of 500 links, which ranked were by the the Rotation-Forest classifiers as links with the highest likelihood  of being restrict by the SPP application users, 470, 455, and 440 links were actually restricted in the Fake profiles datasets, Friends restriction dataset, and All links dataset respectively.

In order to estimate the classifiers' recommendations precision according to the SPP users' point of view, we also calculated the Rotation-Forest classifier average users precision \textit{k}, as described in Section~\ref{sec:construct}, on the different datasets for different values of \textit{k}\footnote{In case of the Friends profiles datasets, we  calculated the average users precision for 355 SPP application users only, which for certain were familiar with alphabetical interface and used it to restrict their friends.  }. 

\begin{figure}[ht]
\begin{center}

\includegraphics[
 width=0.8\textwidth,clip]{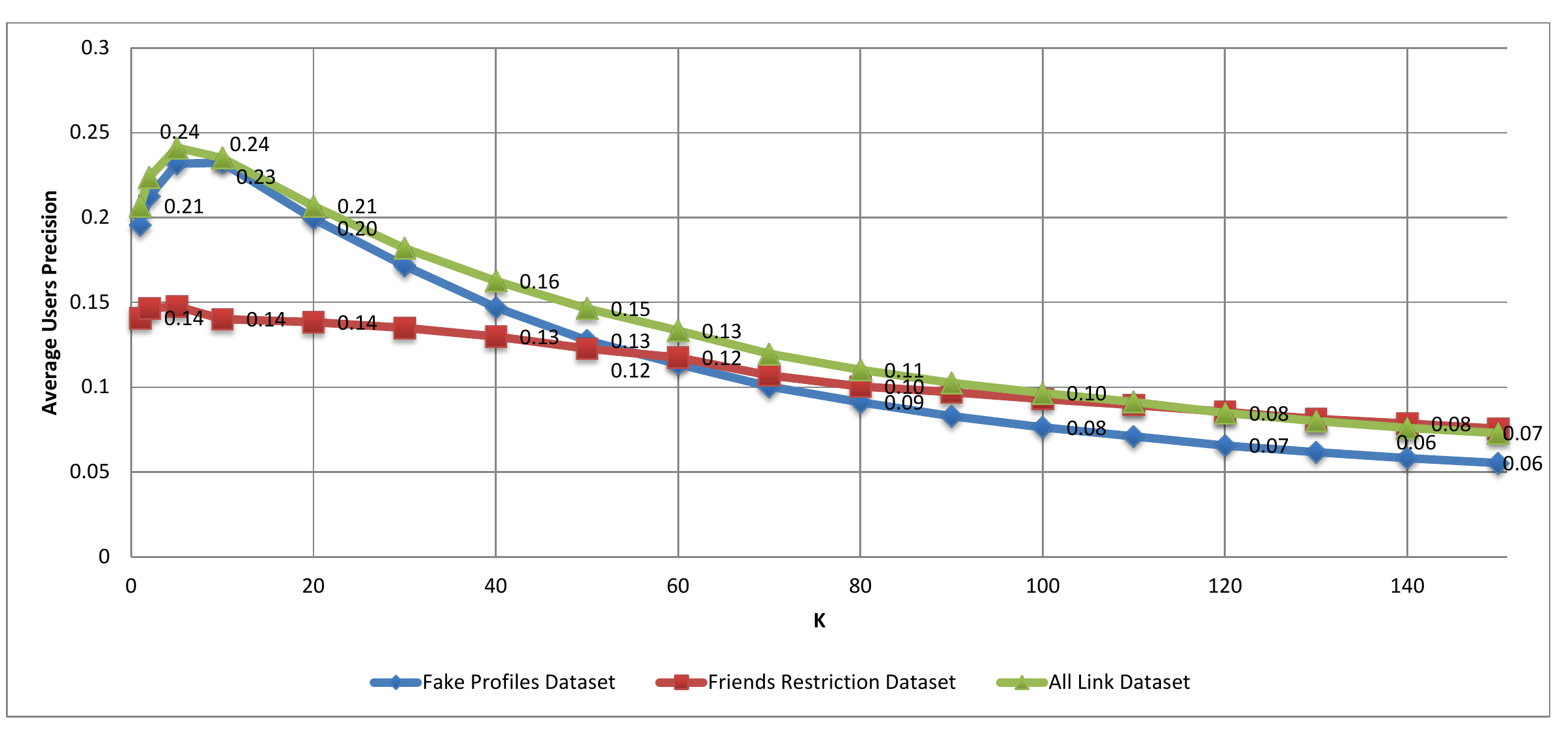}
\end{center}
\caption{Rotation-Forest average users' precision@k - it can been seen that the classifiers' average users precisions at 20 were  21\%, 20\%, and 14\% for the All links dataset, Fake profiles datasets, and Friends restriction dataset respectively.}
\label{fig:rf_avg_precision_k}
\end{figure}

The results of this precision at \textit{k} evaluation are presented in Figure~\ref{fig:rf_avg_precision_k}.
It can be noticed that the Rotation-Forest classifiers presented precision at 10 of 24\%, 23\%, and 14\% for the All links dataset, Fake profiles datasets, and Friends restriction dataset respectively. 
The Rotation-Forest classifier's results on the All links dataset indicated that on average 2.4 of the users' friends, which received the top ten highest probabilities of being restricted among all the friends of each user were actually restricted.  However, the
Rotation-Forest classifier's results on the Restricted profiles dataset indicated that on average only 1.4 of the users' friends, which received the top ten highest probabilities of being restricted among all the friends of each user had actually been restricted.

To obtain an indication of the usefulness of the various features, we also calculated the different features 
importance using WEKA's information gain attribute selection algorithm (see Table~\ref{table:infogain}).
According to the information gain selection algorithm  the top two most useful features on all three datasets were the \textit{Common-Friends }feature and the \textit{Jaccard's-Coefficient} feature. Furthermore, according to the results there are differences between the features scores in the different datasets. For example, the \textit{Common-Groups-Ratio} feature  received a value of 0.113 in the Fake-profile dataset and   a value of only 0.004 in the Friends-restriction dataset, and the \textit{Is-Friend-Profile-Private} received a value of 0.056 in the Friends-restriction dataset and  a value of only 0.0002 in the All links dataset.

\begin{table}[htb]
\caption{Information Gain Values of Different Features for Different Datasets\label{table:infogain}}{ 
\centering
\includegraphics[width=\textwidth]{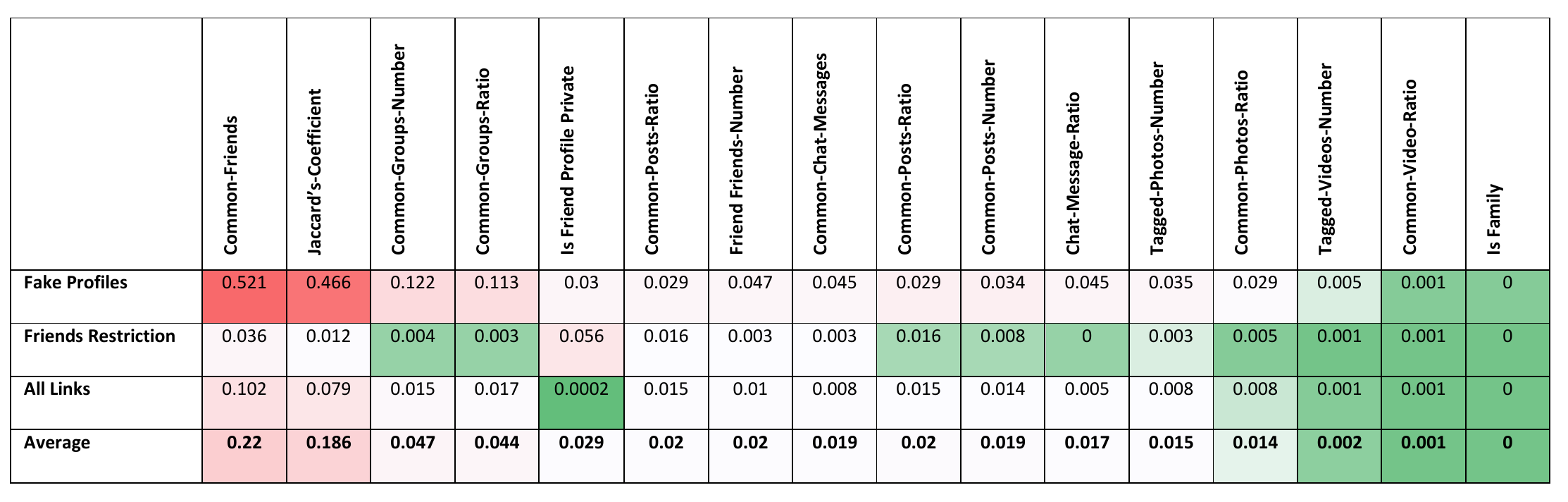}
}
\end{table}

\subsection{Add-on Results}
\label{sec:results_addon}
The SPP Firefox Add-on was downloaded more than 1,676\footnote{The SPP Add-on was available for download from several locations, such as the Firefox Add-ons website and the PrivacyProtector.net  website.  Due to the fact that not all locations store the number of downloads, we can only estimate the number of downloads according to our HTTP Server logs.} 
times between the 27th of June, 2012 and the 10th of November, 2012.  
During that time we succeeded in collecting data with the number of installed Facebook applications from 1,676 different Facebook users. 
This data was collected on 21,524 different occasions.
Furthermore, we also succeeded in collecting SPP users' privacy settings of at least 67 Facebook users on 129 different occasions\footnote{Due to the fact that not all SPP users opened their Facebook privacy settings during this time period, and probably due to problems in parsing the different Facebook privacy settings page layouts, we succeeded in collecting the SPP users' privacy settings for only a limited number of users.
 }.
 
\begin{figure}[ht]
\begin{center}

\includegraphics[
 width=0.8\textwidth,clip]{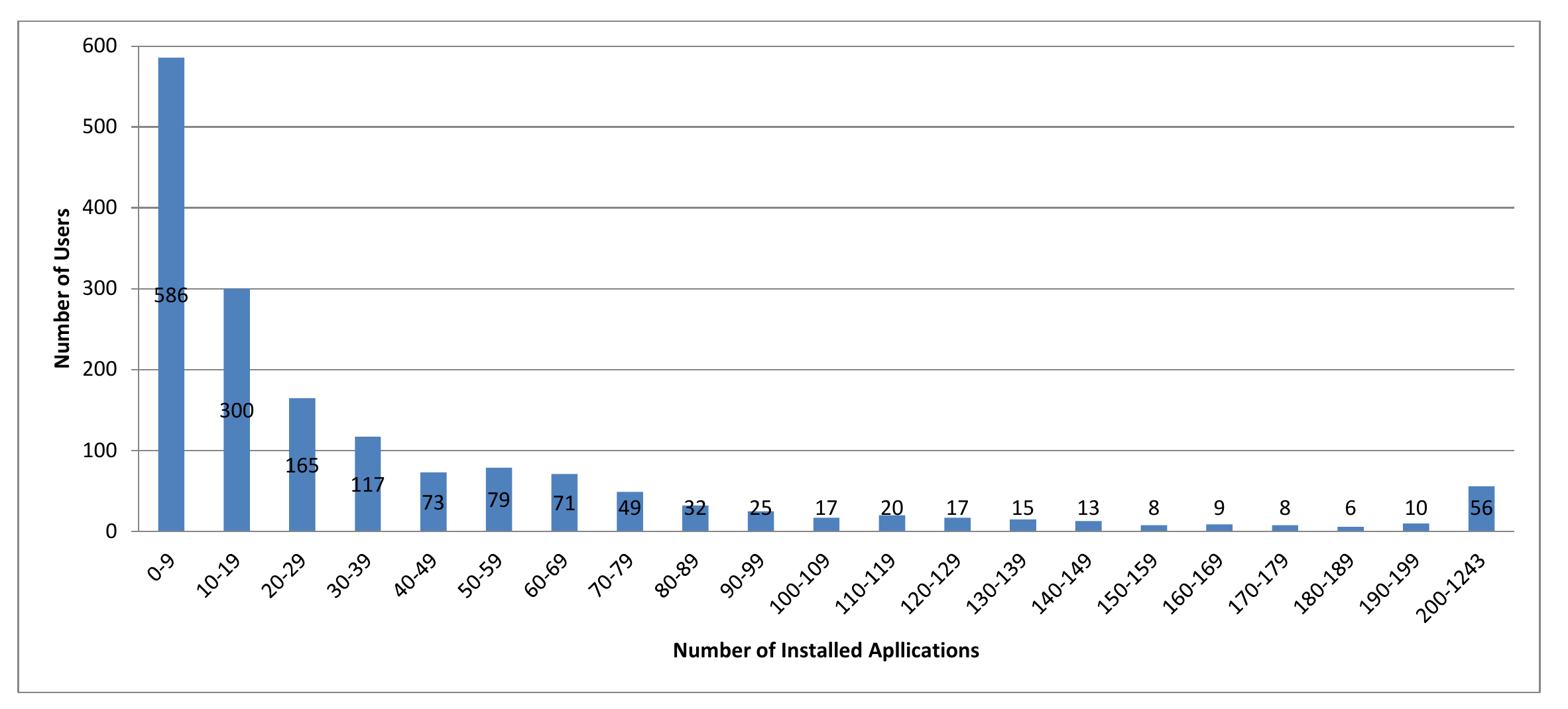}
\end{center}
\caption{Distribution of The Number of Installed Facebook Applications - it can be noted that 10.68\% of the users have more than 100 applications installed on their profiles.}
\label{app_dis}
\end{figure}

By analyzing the collected applications data we discovered that the number of Facebook applications installed on users' profiles, at the time they initially installed our Add-on, ranged from one installed application to 1,243 installed applications, with an average of 42.266 applications per user. 
Moreover, according to the installed applications distribution, we can observe that about 34.96\% of the users have less than ten applications installed on their profiles. However, 30.31\% of the users have at least 40 installed applications and 10.68\% have more than 100 applications installed on their profiles  (see Figure~\ref{app_dis}).

\begin{figure}[ht]
\begin{center}

\includegraphics[
 width=0.8\textwidth,clip]{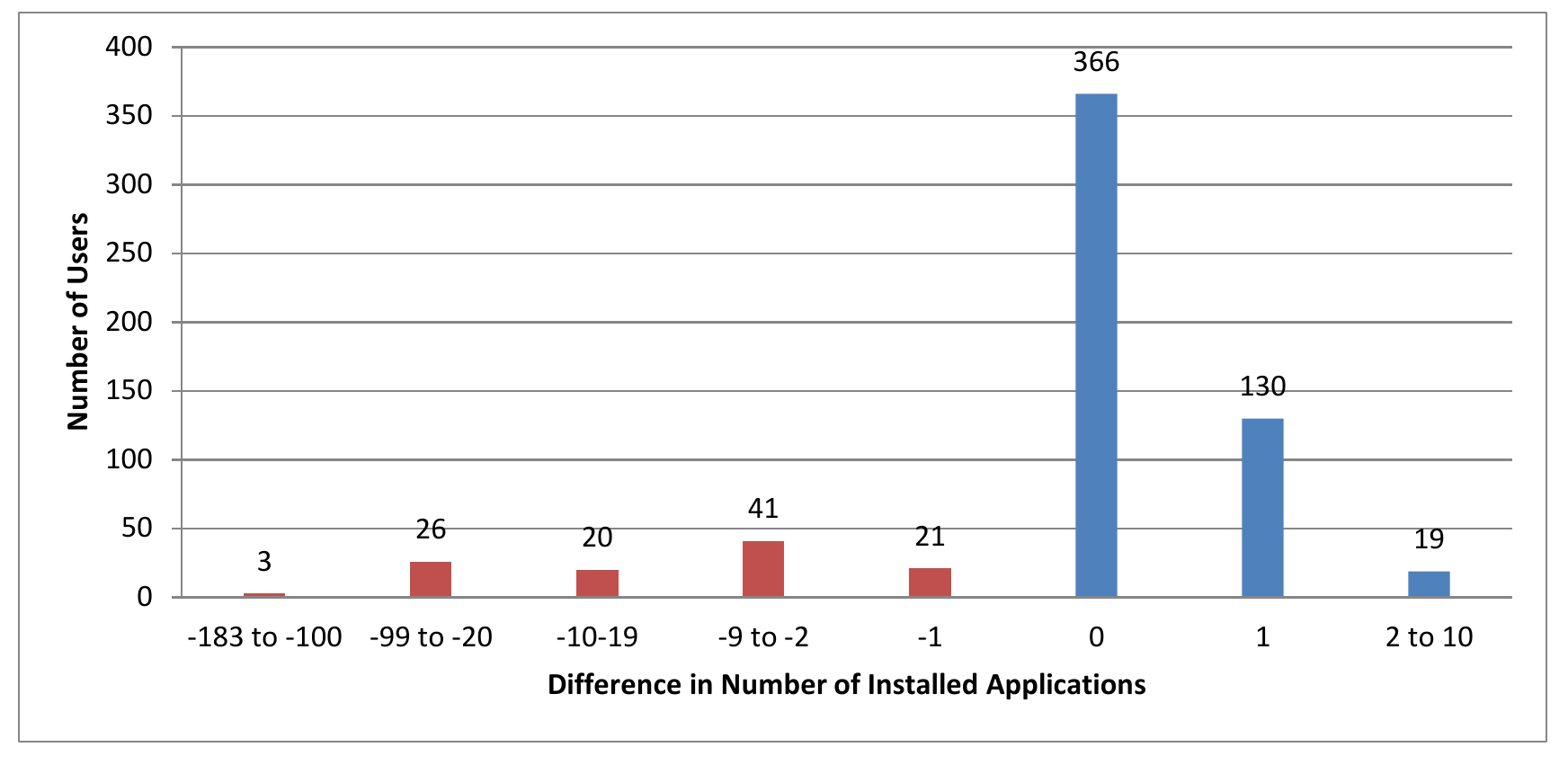}
\end{center}
\caption{Distribution of the Difference in Installed Applications Number Day After the Add-on Installation.}
\label{app_diff}
\end{figure}

\begin{figure}[ht]
\begin{center}

\includegraphics[
 width=0.8\textwidth,clip]{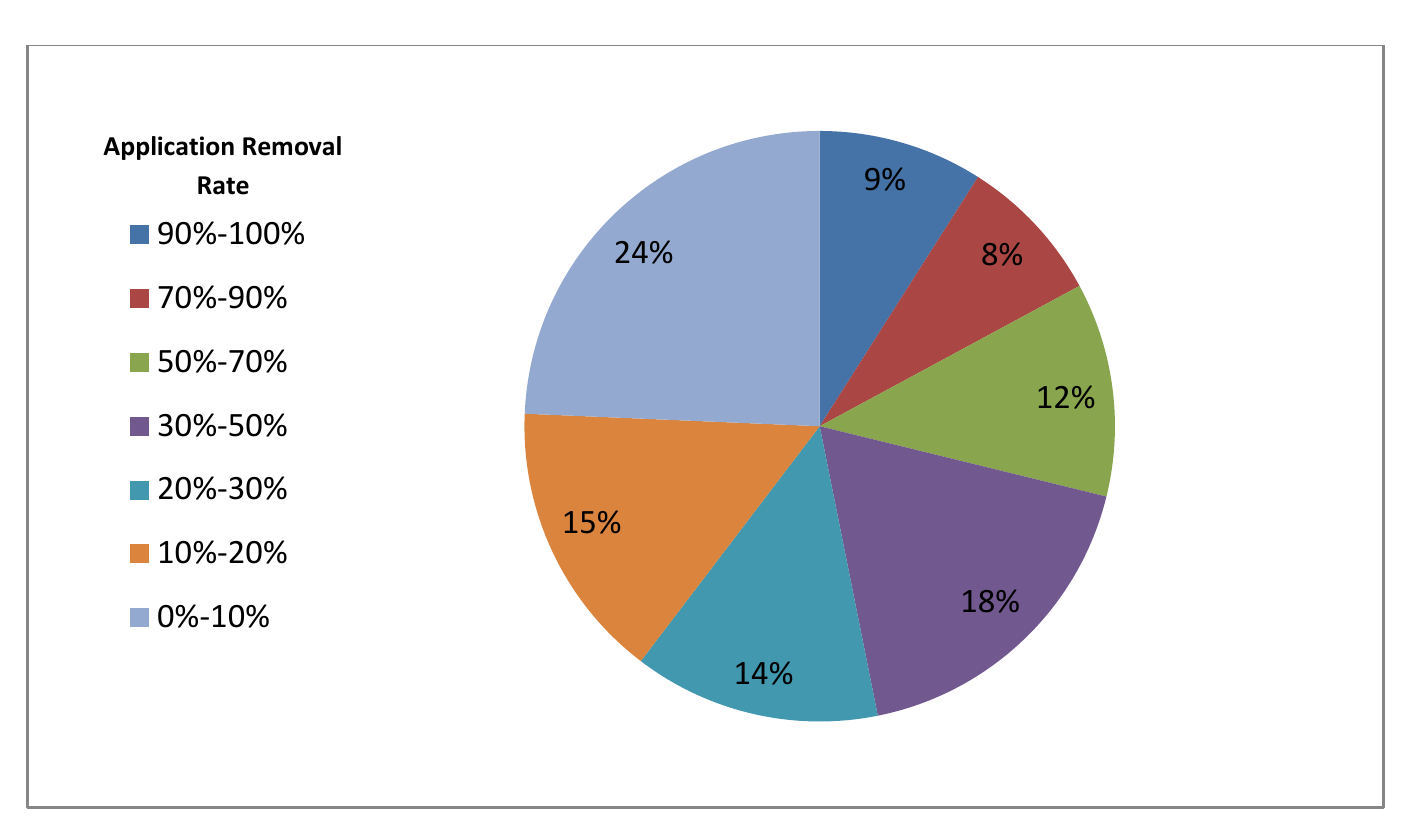}
\end{center}
\caption{Removed Application Percentage - every slice of the pie represent the percent of applications removed by a percent of users. For example, 9\% of the Add-on users removed between 90\%-100\% of all applications installed on their profiles. It can be noted that about 29\% of all the users have removed more than 50\% of all applications installed on their profiles.}
\label{app_ratio}
\end{figure}

In addition to calculating the statistics on the number of installed Facebook applications, we also tested if the SPP users had used the Add-on to remove part of their installed Facebook applications. 
In order to identify if a user has removed the Facebook applications using the Add-on, we check what the user's installed applications numbers up to a day after the Add-on is initially installed.
Our Add-on succeeded to collect the data of 626 users a day after the  Add-on is initially installed.
Out of these 626 users 111 (17.73\%), had removed 1,792 applications, while 149 (23.8\%), users added 192 applications, and 366 (58.47\%) users did not add or remove any applications (see Figure~\ref{app_diff}). 
A closer look at the application removal data reveals that on average each user from the 111 users removed 34.7\% of all installed applications and 32 (28.8\%), users had removed at least 50\% of all their installed applications (see Figure~\ref{app_ratio}) .

If we look at the overall time period of our experiments, from the 27th of June, 2012 to the 10th of November, 2012, we can see that out of 1,676 users 335  (19.99\%), users decreased the number of installed applications on their profiles. These users had removed 5,537 applications with an average of 16.52 application removals per user and a median of seven.  

In addition to checking how many applications were removed, we also checked how many new applications were installed to each of the Add-on users. To achieve this goal, we first focused on the group of Add-on users, which to the best of our knowledge had more applications installed on their profile at the end of the 10th of November, 2012 than in the day they first installed the Add-on. For these user groups, we calculated how many applications were added to their profiles on an average each week. We discovered that out of 1,676 users, 389 (23.2\%), users increased the number of installed applications on their profile ranging from 0.05 to 107.33 average of new application installations per day (with a median of 0.636 and an average of 1.91). 

We also analyzed the distribution of the privacy-settings collected from the 67 unique Add-on users (see Table~\ref{table:privacy}). It can be noticed that 74.62\%  of the users set their default privacy settings to be exposed to everyone. 
Moreover, according to the users privacy settings it can be noticed that almost all the user information, except  \textit{Tag-Suggestions}, is exposed to the friends of the user. In addition, we also tested how many Add-on users changed their privacy settings during this time period, and discovered that according to our logs 14 (20.9\%) Add-on users changed their privacy settings. 
However, after a short while the majority of these fourteen users returned to their old, less restricted privacy settings.

\begin{table}[ht]
\caption{Add-on Unique Users Privacy Settings Distribution\label{table:privacy}}{
\begin{tabular}{ |c |c|  c| c| c| c| c| c|  } 
\hline 
Users 	 &Default & Lookup & Share & Send & Receive Friend & Tag & View\\  
Settings &Privacy &  &Address  & Messages & Req. & Sugg. & Birthday \\
		&Settings & &  & & &  & \\
\hline 
Public/ 	&  74.62\% &  74.62\% &  - & 74.63\% & - &  - & 11.94\% \\
Everyone 	& & & & & & &  \\
\hline
Friends 	& 23.88\% & 25.38\%   & 100\% & 25.27\%& - & 52.87\% & -\\
\hline
Friends &  - & - & - &  - & 86.76\% & -& 88.06\% \\
of Friends & & & & & & &  \\
\hline
No one &  - & -&- & - & - & 23.88\% & \\
\hline

\end{tabular}
}
\end{table}

\section{Discussion}
\label{sec:discussion}

By analyzing the results presented in Section~\ref{results}, we can notice the following:

First, we notice that the initial SPP application results presented relatively good performances. Although we defined the Connection-Strength heuristic to  be quite simple; it presented remarkable precision, where on average 31.1\% of the users' friends, which were ranked in first place were actually restricted (see Figure~\ref{fig:ranking_rates}). Furthermore, the heuristic also presented an average users precision of 33.6\% at 1, an average users precision of 27\% at 10, and an average users precision of 11\% at 100 (see Figure~\ref{fig:cs_avg}). 
However, the Connection-Strength heuristic was not able to present a generic method, with high true-positive rates and low false-positive rates, for recommending the restriction of links. For example, only 10\% of the SPP users' friends who received a Connection-Strength with a value of 3 were restricted (see Figure~\ref{fig:cs_stat}). 

Second,  among all tested machine learning algorithms the Rotation-Forest classifiers performed best on all datasets, with especially good results for AUC of 0.948 and a false-positive rate of 15.8\% on the Fake profiles dataset (see Table~\ref{table:10_folds}). 

Third, according to the results, the Rotation-Forest classifiers' average users precision at 1 on the All links dataset  datasets and on the Fake-profiles it was 21\% and 20\%, respectively. These results were worse than the results presented by the Connection-Strength heuristic, which presented an average users precision at 1 of 34\%. However, the classifiers average users precision at \textit{k} for higher \textit{k} values was nearly the same as the Connection-Strength heuristic's precision at \textit{k}. For example, Connection-Strength heuristic average users precision at 20 was 22\%, while the Rotation-Forest classifiers' average users precision at 20 was 21\% on All links dataset and 20\% on Fake-profiles datasets (see Figures~\ref{fig:cs_avg} and~\ref{fig:rf_avg_precision_k}). 
Nevertheless, using Rotation-Forest classifiers has many advantages, which the Connection-Strength heuristic does not have, such as presenting a generic model for links restriction recommendation, and  presenting the restriction probability for each link in the network without the need to compare each link to other links of the same user as is done in the case of the Connection-Strength heuristic.

Fourth, in contrast to the other classifiers,  the  classifier which was constructed from the Restricted friends dataset using the Rotation-Forest algorithm presented a relatively low average users precisions at 1 of 14\% (see Figure~\ref{fig:rf_avg_precision_k}). However, this precision is significantly better than the precision obtained by random guessing, which stands on 4.25\%, in the Friends restriction dataset. 
We assume that this classifier presented a relatively low performance because the restricted friends in this dataset were  mainly real friends, which the SPP users chose to restrict for different reasons. We assume that these reasons cannot be inferred from the features we extracted in the SPP's initial version. 

Five, when the Rotation-Forest classifiers were evaluated on the general scenario of predicting which links to restrict among all users' links, the classifiers presented very high precision rates. For example, the Rotation-Forest classifier, which was constructed from Fake profiles dataset links presented 91\% precision at 100 and 94\% precision at 500 (see Figure~\ref{fig:precision_k}). 
Moreover, the Rotation-Forest classifier, which was constructed from the Friends restriction dataset presented impressive precision at 100 of 98\%. These results indicate that the classifiers can be used not only by the social network users, but also by the online social network administrator in order to identify fake profiles among all profiles on the network.

Six,  according to the information gain results we can conclude that on all datasets the most useful features were the \textit{Common-friends} feature and the \textit{Jaccard's-Coefficent }feature (see Table~\ref{table:infogain}). Additionally, the \textit{Is-Friend-Profile-Private} was found to be very useful in the case of the Friends restriction dataset indicating that friends, which have their profile set to be private have a higher likelihood of being restricted. Moreover, according to these results it is noticeable that the \textit{Is Family}, and the \textit{Tagged-Video-Number} features were not so useful. Removing the extraction of these features in future versions can assist in improving the SPP application run time without significantly affecting the results.

Seven, according to the applications statistic results it can be noted that many Facebook users installed many applications on their Facebook accounts, which can jeopardize their privacy. According to our results out of 1,676 examined Facebook users, 30.31\% of the users had at least forty installed applications and 10.68\% of the users had more than a hundred installed applications (See Figure~\ref{app_dis}). 
Moreover, according to our results, out of the 111 users, which used the SPP Add-on for application removal, 28.2\% removed more than 50\% of all their installed applications a day after they installed the SPP Add-on (see Figure~\ref{app_ratio}). These results indicate that in many cases the installed applications are unwanted or unneeded applications. Furthermore, our results also uncovered an alarming phenomenon; namely, many Facebook users install new applications weekly. According to our results 389 out of 1,676 users had increased the number of installed Facebook applications on their profiles with an average number of 1.91 new application installations per week. 

Eight, according to the collected users' privacy statistics we can see that almost all of the examined users information is available to friends, leaving the users' information exposed to fake friends (See Table~\ref{table:privacy}). In  addition, the vast majority of examined users also set their ``Default Privacy Setting'' to be accessed by everyone. This result indicates that many users do not protect their personal information and leave it exposed to the public's view.

Lastly, according to the overall results we can see that SPP software has assisted its users in protecting their privacy both by restricting friends and by removing unwanted Facebook applications. However, the SPP software did not succeeded in assisting users to improve their privacy settings in most cases.

\section{Conclusions}
\label{sec:conclusions}

In this study, we presented the SPP software, which aims to better protect user's privacy in Facebook.
We presented in detail the general architecture of the SPP software (see Section~\ref{architecture}). 
According to this architecture, the SPP software can be divided into three layers of protection. The first layer helps to restrict a user's friend's access to personal information. The second layer helps to identify and warn the user about installed Facebook applications, which can violate the user's privacy. The third layer helps the user to adjust their privacy settings with one click.  
According to this software architecture the heart of the SPP software lays in the function, which is responsible for recommending to each user, which friends to restrict. This function can be a simple heuristic or a more complicated machine learning classifier.

In the initial version of the SPP software we chose to implement the Connection-Strength heuristic, which was responsible for recommending to each user, which of his friends to restrict (see Section~\ref{spp_features}). 
By using the software interfaces and the Connection-Strength heuristic recommendations, 527 out of 3,017 software users had restricted 
9,005 friends in less than four months. 
According to the results of the Connection-Strength heuristic  we can conclude that the Connection-Strength heuristic presented the users with a relatively remarkable recommendation. By using these recommendations the SPP users' had restricted 30.87\% of their friends, which appeared 
in the first position in the application's Restriction interface (see Figure~\ref{fig:ranking_rates}).
However, the Connection-Strength did not provide a general method of identifying which of the users' links need to be restricted.

To create general link restriction recommendation methods we chose to use a supervised learning approach. 
By using the unique data, which was created by the initial SPP version, we created three types of datasets for different friend restriction scenarios(see Section~\ref{sec:methods}). We used these datasets to construct and to compare different machine learning algorithms  to identify which algorithm can provide the best results (see Section~\ref{sec:construct}). 
We discovered that the Rotation-Forest algorithm presented the best AUC and false-positive rates results on all three datasets (see Table~\ref{table:10_folds}). We then showed that the Rotation-Forest classifiers had been created from the Fake profiles dataset and the All links dataset presented good average users precision at \textit{k} results (see Figure~\ref{fig:rf_avg_precision_k}).
Furthermore, we demonstrated that these classifiers can provide Facebook administrators with a method, which can assist them in identifying fake profiles among all users in the network (see Figure~\ref{fig:precision_k}). 
However, according to our evaluations these classifiers suffer from relatively high false-positive rates. We believe that these false-positive rates can be considerably reduced if the network administrator uses the classifiers to evaluate several links, instead of one link only, for each suspicious profile, which the classifiers had marked off as being a fake profile. We hope to verify this assumption in future research.

In this study, we also collected statistics form 1,676 different Facebook users on the number of applications installed on their Facebook profiles during different time periods. According to these statistics we discovered that many Facebook users have an alarming number of applications installed on their profiles, where 30.31\% of the users had at least forty installed applications (See Figure~\ref{app_dis}). In addition, our statistical analysis also showed that many users continued to install new applications with an average of 1.91 new applications every week. Fortunately, according to our results, we discovered that at least 111 SPP Add-on users had used the SPP Add-on  to improve their privacy and removed at least 1,792 applications (see Figure~\ref{app_diff}). These results indicate that by making users more aware of the existence of installed applications we can assist in reducing the number of the installed applications, and may decrease the exposure of users' personal information to third party companies.
Furthermore, in this study we also collected statistics on the privacy settings of 67 unique Facebook users. 
According to these privacy statistics we can conclude that a majority of the users expose their private information to friends, and in many cases even to the public. Once again, these statistics sharply demonstrate how exposed Facebook users information can be to both fake profile attacks and  third party Facebook applications. 

In the future, we hope to continue our study and provide an updated version of the SPP Add-on, which will be able to support more web browsers, such as Chrome and Internet Explorer. In addition, we plan to remove the extraction of less useful features, like the \textit{Tagged-Video-Number} feature,  and through this improve the SPP application performance. 
We hope that these improvements will assist SPP users to restrict more fake profiles and through this increase the size of our classifiers' training set.

We also believe that this study has several future research directions, which can improve the identification of fake profiles in online social networks.
A possible direction is to extract more complicated topological features, such as the number of communities of each user, and use them to construct better classifiers with lower false-positive rates. In our previous study~\cite{fire2012strangers}, we demonstrated that these type of features can assist in identifying fake profiles. 
Another possible direction, which can be used to improve the classifiers performances is to construct the classifiers by using oversampling techniques, like SMOTE~\cite{chawla2011smote},  to deal with the dataset imbalance issue instead of the under sampling techniques we used in this study. We also hope to test the constructed classifiers performance on different online social networks, such as Google+ and Twitter.
Another future direction we want to examine is the usage of the SPP software as an educating tool. 
We also hope to examine if users utilizing the SPP software to restrict friends and remove applications became more aware of their privacy, and as result tended to accept less friend requests and installed fewer applications.
In future studies, we also hope to perform a deeper analysis on the unique datasets we obtained with the SPP software and extract different insights on connections between Facebook users. We also hope to test the developed algorithms to improve users' security and privacy  in other online social networks.

\section{Availability}
The Social Privacy Protector and parts of its source code are available for download from \url{http://www.socialprotector.net}.
The Friend Analyzer Facebook application is available to download from \url{https://apps.facebook.com/friend_analyzer_app}. A video with detailed explanations on how to use the SPP application is available in \url{http://www.youtube.com/watch?v=Uf0LQsP4sSs}



\bibliographystyle{abbrv}
\bibliography{privacyprotector}


\end{document}